# A new generation of science overlay maps with an application to the history of biosystematics

Sándor Soós
Dept. Science Policy & Scientometrics,
Library and Information Centre of the Hungarian Academy of Sciences
Arany János. u. 1. H-1051 Budapest, Hungary
E-mail: soos.sandor@konyvtar.mta.hu

**Abstract.** The paper proposes a text-mining based analytical framework aiming at the cognitive organization of complex scientific discourses. The approach is based on models recently developed in science mapping, being a generalization of the so-called *Science Overlay Mapping* methodology, referred to as *Topic Overlay Mapping* (TOM). It is shown that via applications of TOM in visualization, document clustering, time series analysis etc. the in-depth exploration and even the measurement of cognitive complexity and its dynamics is feasible for scientific domains. As a use case, an empirical study is presented into the discovery of a long-standing complex, interdisciplinary discourse, the debate on the species concept in biosystematics.

## Introduction

The empirical study of the cognitive structure and dynamics of science, called science mapping, has, in recent years, been a rapidly evolving field of research within information science. A succesful science mapping approach based on the (aggregated) communication patterns of scholarly journals (Leydesdorff and Rafols 2009), that is, a complex network of Web of Science Subject Categories triggered a complete analytical framework, *viz.* Science Overlay Mapping (*ScOM;* Rafols et al 2010) . The core idea is that any kind of S&T actor (individual, group, institution, country, region etc.) can be modelled by overlaying its research profile on this global science map. Beyond "x-raying" research activity, various structural properties of research profiles can be studied and even quantified based on the underlying science map, such as patterns of inter- and multidisciplinarity—this is why a toolbox of ScOM-based measurements has been developed within interdisciplinarity research or "IDR" (Rafols and Meyer 2010).

The main aim of the work presented below is to elaborate a multi-purpose framework, inspired by Science Overlay Mapping, aiming at the analysis of the cognitive (thematic or topical) organization of scholarly discourses. The backbone of our proposed framework is identical to that of ScOM: (1) Identify the main topics in a scholarly corpus, (2) Draw a global "discourse" map showing its internal cognitive organization by the interrelations of topics and (3) build an analytical toolbox to overlay any parts of the corpus on the global map, and quantify its cognitive patterns by exploiting the undrelying map. The resulting analytical framework (Topic Overlay Mapping, hereafter: TOM) naturally lends itself to three main applications: (1) the measurement of cognitive complexity of papers or paper sets (2) the comparison and clustering of papers based on their overlay maps, and (3) the measurement and visualization of the cognitive dynamics and development of the whole body of literature under study.

As an actual use case for demonstrating the rich analytical potential of TOM, we present an empirical study addressing a historical debate at the intersection of biosystematics, evolutionary biology and the philosophy of science, namely, the so-called species problem. A TOM-based trend analysis is made on the collected corpus, including document clustering (to reveal main lines of the discourse). Instead of the usual practice in text mining research, whereby classification methods are evaluated against pre-established benchmarks on synthetic datasets, we rather interested in science mapping performance, that is, in how effectively the resulting clusters reveal (also, as opposed to VSM-based clustering) visible complexities of the issue.

## Methods

Science overlay maps (ScOMs), as introduced by Rafols and Leydesdorff (Rafols et al. 2010), are based on a complex network of research areas (WoS journal categories, in this case), drawn from their respective proximities in terms of referencing behavior (citation patterns)—called the *basemap* (of science). Such a global map of science is then used to represent research profiles from any set of WoS-indexed papers, showing how those papers are distributed over the basemap—resulting in the *overlay map*. Upon this model, a rich analytical toolbox can be developed to quantify structural properties of a research profile (cf. Rafols and Meyer 2010, Soós and Kampis 2011, 2012). In close analogy with this approach (or as an extension of the method), our framework (TOMs) consists in the following modules:

### *The construction of Topic Overlay Maps*

The central concept in our model is the Topic Overlay Map (TOM), designed to represent the position of any aggregate of papers within a rich cognitive map of a scholarly discourse. Given a corpus (or a set of bibliographic metadata) associated with a scientific discourse, a TOM is created in four steps:

(1) A term-proximity (weighted) graph is obtained from the whole corpus (based on the joint distribution of textual descriptors, e.g. keywords, within the corpus).

(2) The graph is clustered (with an appropriate community detection algorithm) into cohesive term sets, as proxies for main topics of the discourse.

(3) As the key step, upon the preestablished cluster structure, a new graph is constructed that captures the relationship of clusters emerged from the simple term network. Formally, this graph is a proximity network of clusters (as nodes), based on the connectedness of their respective elements (terms) in the underlying term graph. As this new structure is supposed to formalize the cognitive organization of the discourse by a weighted proximity network of topics (proximities expressing topical interrelations), it serves as the *basemap* in our model.

(4) The final step consists of overlaying a selected set of documents (belonging to the corpus) on the basemap. This can be done by remodelling documents in terms of the clusters of the basemap. Each paper or paper set is characterizable via the distribution of its textual descriptors over topics (the cluster set). This distribution can be visualized on the basemap (cf. subsection $C$), or used quantitatively in map-based structural measures (cf. $B$), and is called (along with the map properties) the *overlay map* for the paper set.

*Quantifying structures of cognitive organization*

A peculiar feature of ScOMs for the study of research profiles is that, contrary to simple distributions which allow for measurements of the *variety* and *balance* of constituent research fields, science maps also provide indicators for the *disparity* of the profile, i.e. the cognitive distances between research areas involved in the profile. It is a benefit inherited by Topic Overlay Maps as well, which can be exploited by applying structural map-based measures originally proposed to quantify the diversity (multidisciplinarity) and cohesion (interdisciplinarity) of research profiles. We developed our framework to involve the following applications:

(1) *Quantifying cognitive (topical) complexity of publication(s)*. The so-called Stirling index (Stirling 2007) is applicable to a topic overlay map (cf. Rafols and Meyer 2010), measuring both the topical composition and its cognitive scope for publication(s), indicating the underlying topical complexity. The cognitive scope is based on the distances of constituent topics within the basemap.

(2) *Similarity of overlays and document clustering.* The generalization of the classical Cosine similarity measure, called the Proximity Weighted Cosine Similarity (Zhou et al. 2012) is applicable to compare any two topic overlays. Beyond the similarity of papers in topic composition, it is also sensitive to the cognitive proximity of any two papers (overlays). This similarity measurement lends itself to assist a document clustering that results in paper groups reflecting complex relations or "functional overlaps" between topics rather than their simple combinations.

*Visualizing the cognitive structure and dynamics*

A third application of the framework of TOMs lies in its capacity for visualizing the cognitive organization of discourses, along with their dynamics. Topic Overlays are, in the primary sense, visualizations of a particular topical portfolio (of a paper set) upon the basemap, that reveals the relatedness of topics as well. The visualization consists of customizing the basemap via setting node sizes to express the contribution of each topic (node) to the portfolio, node size being proportionate to the degree of topic contribution. In such a way, the evolution of a discourse can also be set out visually by overlaying the underlying corpus partitioned into consecutive time periods: the dynamics of topics and their interactions as well as their development can be tracked throughout the history of the discourse.

## An empirical study : the interdisciplinary debate on the species concept

In order to test and demonstrate the capacity of the proposed method, we applied it in an attempt to reconstruct the historical development of a rather complex discourse in biology, usually referred to as the Species Problem. The Species Problem can be briefly described as a historical debate on what biological species are, and as the related quest for the appropriate definition of species, or species concept for biology. With a long prehistory, dated as back as to Aristotle and Plato, including Darwin's paradigm-shifting work on the nature of species in the XIX. century, the debate expanded in the early XX. century, mainly due to the rediscovery of Darwin's work, and having it integrated with the early (Mendelian) genetics of the era. The new paradigm has been called the Evolutionary Synthesis. Since the Synthesis, a plethora of theories has emerged on species, resulting in a variety of competing species concepts. According to a comprehensive review of Mayden (1997), no less than 22 species concept (definitions) exhibit themselves in the contemporary literature of the subject.

Given its complexities, the Species Problem was an ideal candidate for a bibliometric analysis of —inter-, or multidisciplinary—knowledge diffusion with the proposed methodology:

(1) The roots of the discourse are centuries-old, while there are several contemporary directions of the debate (and of research) as well (cf. Hull 1988, Ereshefsky 1992).

(2) During its modern history (in the XX. century), many schools of biosystematics contributed to, and competed over the problem, involving—from a data-mining perspective—different topics: theoretical papers as well as empirical ones, the latter focusing on particular subjects of taxonomy (description of taxa). It was of outstanding interest whether the TOM toolkit was capable of identifying the interaction of these schools.

(3) A nonstandard feature of the Species Problem is its complexity in terms of the contributing scholarly fields, or even disciplines. For example, a proper interaction of evolutionary systematics, on one side, and the philosophy of science (of biology), on the other side, had a significant effect on the present state of the debate. It was a good challenge for the proposed method of mapping science dynamics to capture the formation of a truly interdisciplinary research.

*Data.*

To cover a representative corpus of the modern history of the discourse, bibliographic data were harvested from three databases of the Web of Science, namely, the SCI, the SSCI and the A&HCI (Science Citation Index, Social Science Citation Index and the Arts&Humanities Citation Index, respectively). Also in a attempt to avoid the potential exclusion of relevant works from the corpus, data retrieval was based on a topic-related query, that did not put any constraints on the set of fields, journals, authors etc. entering the sample. The query was defined to include all records related (topicwise) to any of the following terms: „species problem", „species definition", „species concept". The sample resulted in n=1605 documents.

*Term graph of textual descriptors*

Implementing step 1–2 of building a basemap for the discourse, a term graph was devised from the bibliographic record. As textual descriptors of papers, author keywords, title words and also keywords extracted from the reference list of papers were selected. To obtain the terms, this set of words was normalized with NLP procedures (stemming etc.) A standard similarity network of the most frequent terms was obtained upon the term–document matrix (via the Cosine similarity). Finally, the resulted graph was subjected to a community detection algorithm sensitive to edge weights (modularity maximization via random walks). The procedure yielded in 14 topics. The term graph along with the detected topics (clusters) is depicted on Fig 1.

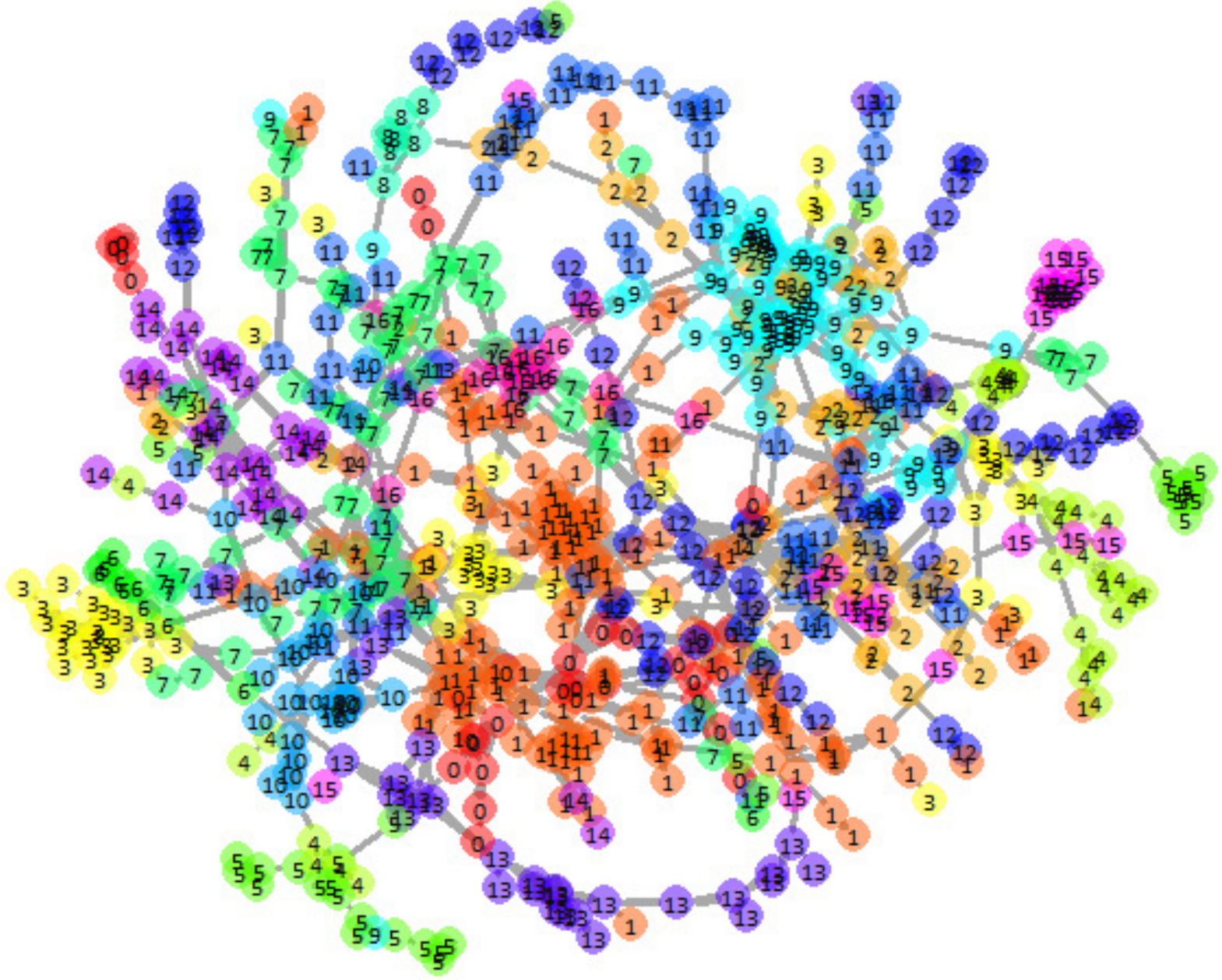

Fig. 1. The term graph and topics (numbered and colored) of the discourse. Only the giant component is shown.

*Basemap of topical relations*

To obtain the basemap (step 3), the relatedness of topics was estimated based on their connection patterns in the term graph. In particular, the “overlap” between two topics *A* and *B* was defined as a weighted average of the weight of edges connecting the elements of *A* and *B*. From running this measure on the term network, a proximity graph of the 14 topics was drawn, yielding the basemap of the discourse for the whole period under study (Fig 2.)

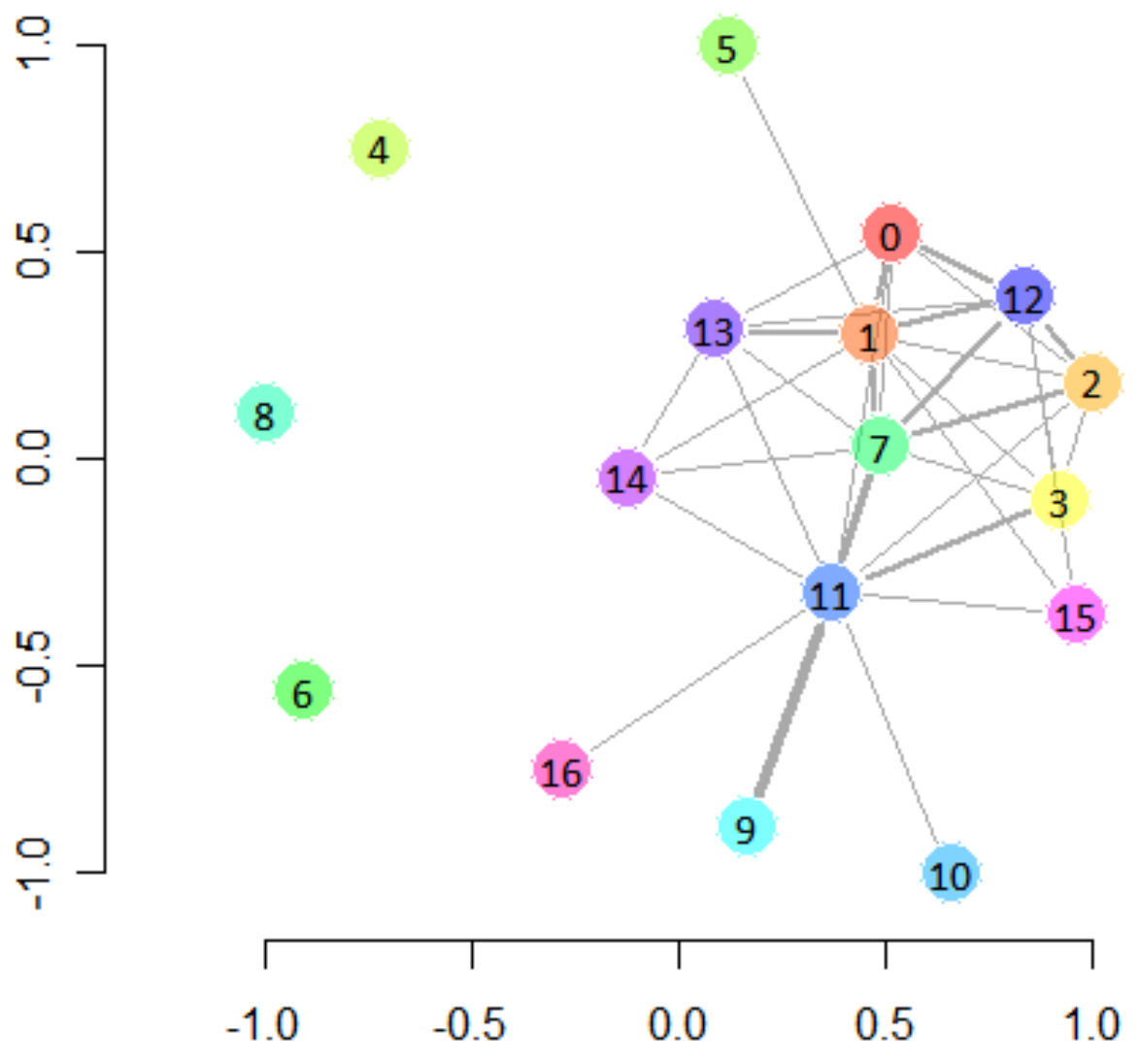

Fig. 2. Topic basemap constructed from the clustered term graph (with the same cluster numbering and coloring). Only the links with strength above a pre-selected threshold are shown.

*Trends in the history of the species debate*

Given the basemap of the selected theme, we used the TOM-based document clustering method, out of the proposed framework, to reveal recent trends in policy-oriented psychological research. Based on the topic overlay map of each paper, after (Zhou et al 2012), we clustered them via their pairwise proximity weighted cosine similarity (cf. section *II.B*, part 2):

$$\varphi(X,Y)=\frac{\varphi_{XY}}{\sqrt{\varphi_{XX}\varphi_{YY}}},$$

whereby $\varphi_{AB}=\sum_{i,j} s_{A(i)B(j)}p_{A(i)}p_{B(j)}$, and

$s_{ij}$ is the proximity of categories A(i) and B(j) within the basemap.

$p_{A(j)}$ and $p_{A(j)}$ are the relative share of topic *i* and *j* in paper *A* and *B*, respectively.

The clustering was achieved by subjecting the resulted similarity matrix to a hierarhic (agglomerative) clustering algorithm (average linkeage clustering).

## Results and discussion

*Comparing the TOM-method against the VSM-based clustering*

As a preliminary assessment of the performance of TOM in our application, we evaluated our TOM-based clustering of the corpus against a classical document clustering method based on the Vector Space Model (VSM). To that end, we also subjected our corpus to a second clustering, whereby paper similarity was established by the standard cosine similarity measure, based on a term–document matrix normalized by the tf–idf method. The grouping of documents was obtained by the same agglomerative clustering algoritm, as in the TOM-based case.

A rather straightforward comparison between the two approaches is to contrast the resulted cluster structures. The dendogram from the TOM-based clustering, and the VSM-based clustering is presented in Fig. 3. and Fig. 4., respectively. The difference of the two dendograms is rather striking: while the TOM-based dendogram conveys a strong internal structuring of the corpus, the VSM-based dendogram mirrors a lack of any proper structure (in fact, this cluster tree was cut at a very high level – that is, extremely low level of relatednes – to even make the graph readable). It suggests that while the standard method was not able to find coherent-enough topics within the discourse, the TOM-based approach could discriminate between topics and even subtopics at a seemingly efficient way.

Since we were primarily interested in the science mapping potential of the TOM toolbox, instead of further quantitative measures, we have focused on the *qualitative* comparison of the two clusterings, concerning the differences of how the corpus is being organized (what kind of trends are being revealed) according to each, respectively. To that end, we have obtained the set of clusters from both dendograms most naturally mirroring their internal structure. Instead of cutting the trees at a pre-defined height, we used the so-called "dynamic tree cut algoritm", that detects the clusters depending on tree shape (http://labs.genetics.ucla.edu/horvath/CoexpressionNetwork/BranchCutting/).

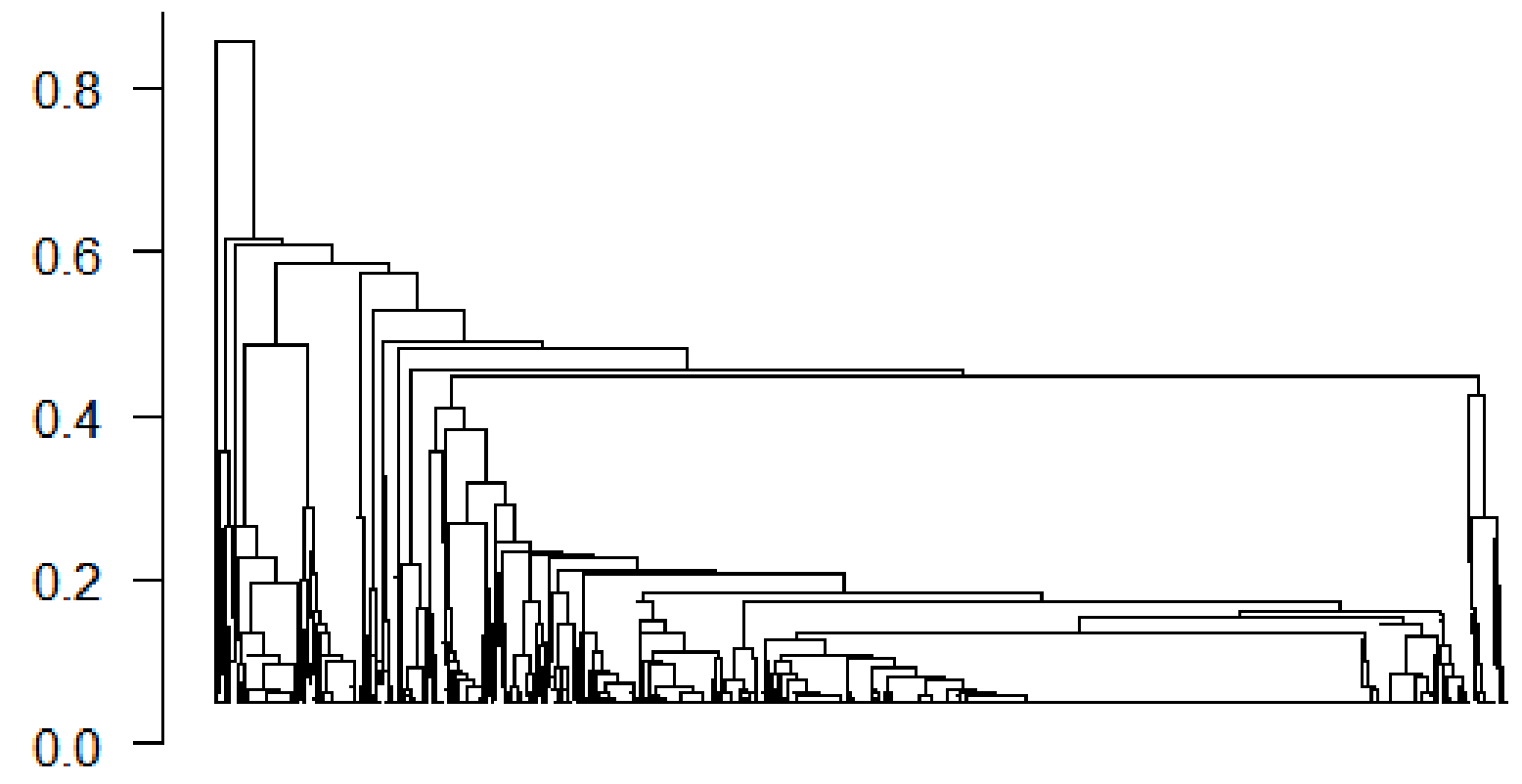

Fig. 3. The cluster structure resulting from the TOM-based clustering of the corpus

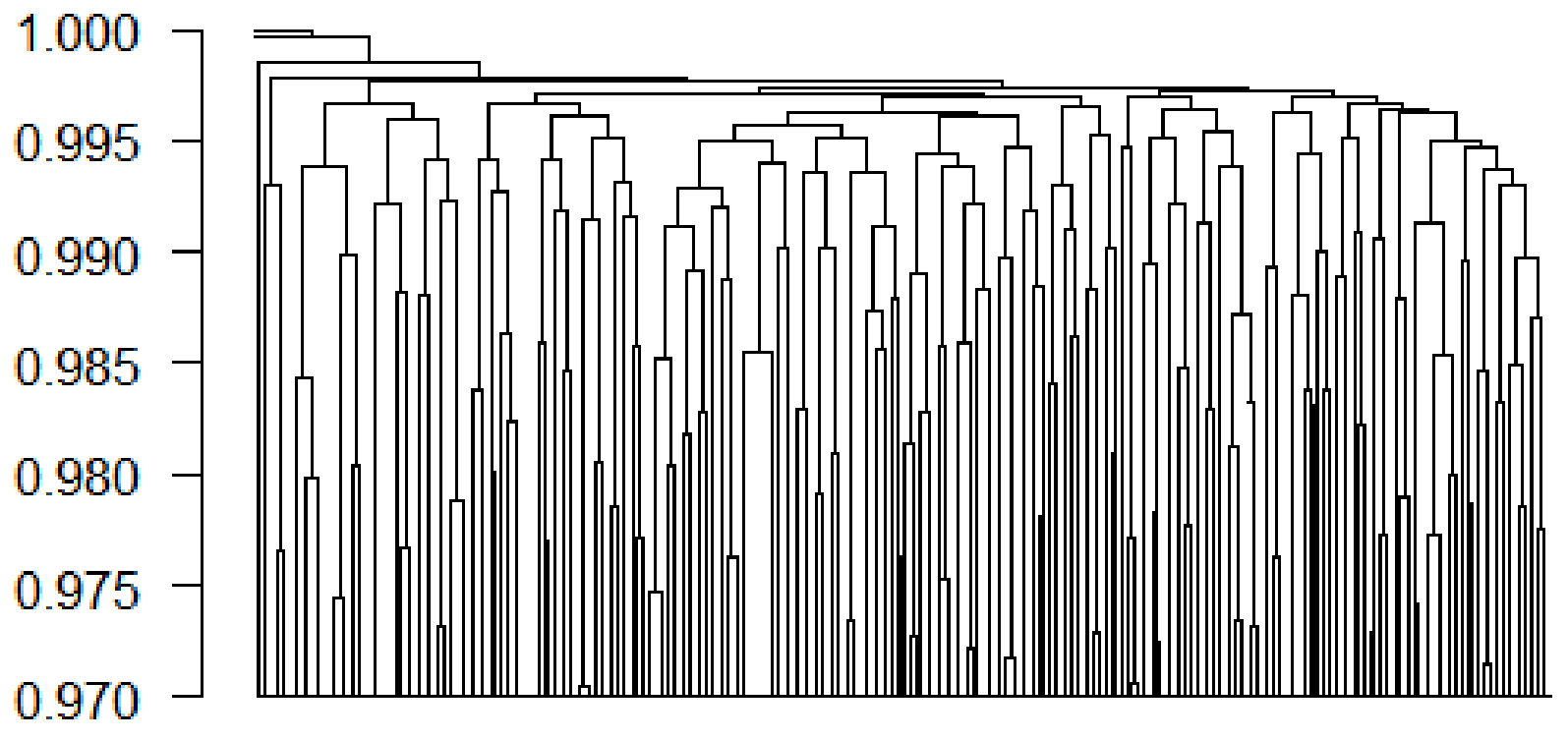

Fig. 4. The cluster structure resulting from the VSM-based clustering of the corpus

The procedure resulted in approx. 20 topical clusters in the VSM-case, and around 50 topical clusters in the TOM-case. To establish a qualitative relation between the two partitions, a keyword-profile was generated for each group in both case, showing the distribution of their most frequent characteristic concepts (author keywords).

The main finding from contrasting TOM-topics and VSM-topics by the overlaps of their keyword profile was consistent with our preliminary expectations. In general, (1) TOM-topics were mostly orgaized around general or "leading" concepts or research subjects indicating many particular lines of investigation that are connected to the general subject, while (2) VSM-topics were more narrow in scope, centered around those particular lines or subtopics. In other words, while VSM.-topics mirrored the "table of content" of the discourse, TOM-topics combined the related "chapters" into single clusters, clearly highlighting the – usually latent – research subject that was central to those chapters.

| clusters | 0 | 1 | 2 | 3 | 4 | 5 | 6 | 7 | 8 | 9 | 10 | 11 | 12 | 13 | 14 | 15 | 16 | 17 | 18 | 19 | 20 | 21 | 22 | 23 |
|---|---|---|---|---|---|---|---|---|---|---|---|---|---|---|---|---|---|---|---|---|---|---|---|---|
| 1 | 14 | 3 | 2 | 9 | 5 | 8 | 2 | 5 | 4 | 2 | 2 | 5 | 0 | 1 | 2 | 1 | 1 | 2 | 2 | 3 | 2 | 1 | 1 | 2 |
| 2 | 17 | 8 | 2 | 1 | 6 | 2 | 0 | 2 | 6 | 0 | 1 | 2 | 0 | 1 | 3 | 0 | 1 | 3 | 1 | 2 | 3 | 3 | 0 | 2 |
| 3 | 10 | 0 | 0 | 2 | 0 | 0 | 1 | 0 | 2 | 15 | 0 | 2 | 0 | 0 | 3 | 1 | 0 | 4 | 0 | 3 | 3 | 1 | 0 | 2 |
| 4 | 24 | 2 | 0 | 0 | 9 | 1 | 0 | 1 | 2 | 0 | 0 | 2 | 0 | 5 | 2 | 0 | 1 | 2 | 0 | 0 | 3 | 3 | 0 | 4 |
| 5 | 12 | 1 | 1 | 0 | 0 | 1 | 28 | 0 | 1 | 0 | 0 | 0 | 33 | 1 | 2 | 1 | 0 | 0 | 1 | 1 | 0 | 0 | 0 | 1 |
| 6 | 27 | 6 | 0 | 0 | 3 | 0 | 1 | 0 | 0 | 0 | 0 | 1 | 0 | 4 | 3 | 0 | 21 | 0 | 1 | 0 | 0 | 3 | 3 | 0 |
| 7 | 26 | 2 | 0 | 0 | 2 | 2 | 3 | 5 | 3 | 0 | 0 | 3 | 2 | 0 | 0 | 2 | 3 | 0 | 17 | 3 | 0 | 6 | 0 | 0 |
| 8 | 11 | 3 | 23 | 0 | 2 | 0 | 0 | 2 | 0 | 0 | 3 | 5 | 2 | 0 | 2 | 10 | 0 | 2 | 6 | 0 | 0 | 2 | 0 | 0 |
| 9 | 20 | 2 | 2 | 0 | 2 | 0 | 3 | 0 | 0 | 17 | 0 | 2 | 0 | 0 | 0 | 0 | 0 | 0 | 0 | 2 | 0 | 0 | 0 | 3 |
| 10 | 9 | 7 | 4 | 2 | 0 | 0 | 2 | 0 | 7 | 0 | 2 | 0 | 2 | 2 | 2 | 0 | 0 | 2 | 0 | 0 | 0 | 2 | 16 | 0 |
| 11 | 12 | 4 | 0 | 2 | 0 | 0 | 2 | 0 | 2 | 0 | 21 | 2 | 0 | 19 | 12 | 2 | 4 | 0 | 0 | 0 | 0 | 0 | 2 | 0 |
| 12 | 31 | 15 | 4 | 0 | 0 | 0 | 0 | 6 | 0 | 0 | 2 | 0 | 6 | 0 | 0 | 2 | 0 | 8 | 0 | 2 | 0 | 0 | 8 | 0 |
| 13 | 16 | 4 | 4 | 0 | 0 | 0 | 2 | 0 | 0 | 0 | 6 | 0 | 0 | 4 | 4 | 0 | 12 | 2 | 4 | 0 | 0 | 0 | 2 | 0 |
| 14 | 11 | 6 | 15 | 0 | 2 | 4 | 2 | 4 | 4 | 0 | 2 | 0 | 0 | 2 | 2 | 13 | 0 | 0 | 2 | 0 | 0 | 2 | 4 | 0 |
| 15 | 13 | 2 | 19 | 0 | 0 | 0 | 0 | 0 | 0 | 0 | 6 | 0 | 0 | 0 | 0 | 23 | 0 | 0 | 2 | 4 | 0 | 0 | 0 | 2 |
| 16 | 29 | 2 | 0 | 0 | 2 | 0 | 2 | 4 | 2 | 0 | 11 | 0 | 0 | 7 | 0 | 0 | 0 | 2 | 0 | 0 | 0 | 11 | 2 | 2 |
| 17 | 29 | 5 | 2 | 5 | 0 | 0 | 2 | 5 | 0 | 0 | 2 | 0 | 0 | 0 | 0 | 0 | 0 | 5 | 0 | 2 | 0 | 2 | 0 | 5 |
| 18 | 3 | 0 | 0 | 0 | 3 | 0 | 0 | 0 | 3 | 12 | 0 | 3 | 0 | 0 | 0 | 0 | 0 | 0 | 0 | 0 | 24 | 0 | 0 | 0 |
| 19 | 18 | 0 | 0 | 0 | 0 | 0 | 0 | 0 | 0 | 0 | 3 | 0 | 0 | 0 | 0 | 0 | 0 | 3 | 0 | 3 | 0 | 0 | 3 | 3 |
| 20 | 12 | 0 | 3 | 0 | 0 | 3 | 3 | 3 | 3 | 0 | 0 | 0 | 0 | 0 | 0 | 0 | 0 | 3 | 0 | 0 | 0 | 0 | 0 | 0 |
| 21 | 16 | 0 | 0 | 0 | 0 | 3 | 0 | 0 | 0 | 6 | 0 | 0 | 0 | 3 | 0 | 0 | 0 | 0 | 0 | 0 | 0 | 0 | 0 | 3 |

| clusters | 24 | 25 | 26 | 27 | 28 | 29 | 30 | 31 | 32 | 33 | 34 | 35 | 36 | 37 | 38 | 39 | 40 | 41 | 42 | 43 | 44 | 45 | 46 |
|---|---|---|---|---|---|---|---|---|---|---|---|---|---|---|---|---|---|---|---|---|---|---|---|
| 1 | 0 | 1 | 1 | 2 | 2 | 0 | 0 | 1 | 1 | 1 | 1 | 0 | 1 | 1 | 0 | 2 | 0 | 1 | 0 | 1 | 0 | 1 | 3 |
| 2 | 1 | 1 | 0 | 4 | 2 | 0 | 7 | 2 | 2 | 5 | 5 | 0 | 0 | 3 | 0 | 2 | 0 | 0 | 0 | 1 | 0 | 0 | 1 |
| 3 | 1 | 5 | 0 | 1 | 3 | 0 | 7 | 1 | 2 | 0 | 0 | 2 | 1 | 0 | 17 | 4 | 0 | 0 | 0 | 2 | 2 | 0 | 0 |
| 4 | 0 | 1 | 3 | 2 | 0 | 0 | 4 | 5 | 1 | 3 | 4 | 1 | 0 | 2 | 0 | 1 | 0 | 0 | 0 | 3 | 0 | 5 | 0 |
| 5 | 2 | 0 | 5 | 1 | 0 | 1 | 0 | 0 | 0 | 0 | 0 | 0 | 0 | 0 | 0 | 0 | 0 | 0 | 1 | 0 | 0 | 2 | 1 |
| 6 | 15 | 1 | 0 | 0 | 1 | 0 | 0 | 3 | 0 | 1 | 0 | 0 | 0 | 1 | 0 | 0 | 1 | 0 | 1 | 0 | 0 | 0 | 0 |
| 7 | 5 | 3 | 3 | 2 | 2 | 0 | 0 | 2 | 0 | 2 | 2 | 0 | 0 | 3 | 0 | 0 | 0 | 0 | 2 | 2 | 0 | 0 | 0 |
| 8 | 2 | 2 | 2 | 3 | 2 | 0 | 0 | 2 | 2 | 0 | 2 | 0 | 0 | 2 | 0 | 3 | 6 | 0 | 0 | 0 | 0 | 0 | 3 |
| 9 | 3 | 3 | 2 | 2 | 0 | 0 | 2 | 0 | 7 | 0 | 0 | 5 | 3 | 0 | 7 | 3 | 0 | 0 | 2 | 3 | 2 | 5 | 0 |
| 10 | 0 | 0 | 14 | 0 | 2 | 2 | 2 | 4 | 0 | 0 | 4 | 0 | 0 | 5 | 0 | 0 | 0 | 0 | 2 | 5 | 0 | 5 | 0 |
| 11 | 0 | 0 | 0 | 0 | 2 | 4 | 0 | 2 | 0 | 0 | 4 | 0 | 2 | 4 | 0 | 0 | 0 | 0 | 0 | 0 | 0 | 0 | 0 |
| 12 | 0 | 0 | 2 | 0 | 0 | 0 | 0 | 0 | 0 | 10 | 0 | 0 | 0 | 0 | 0 | 0 | 0 | 2 | 0 | 0 | 0 | 0 | 4 |
| 13 | 8 | 0 | 0 | 0 | 2 | 4 | 2 | 4 | 0 | 0 | 0 | 0 | 6 | 2 | 0 | 0 | 2 | 0 | 2 | 2 | 2 | 2 | 4 |
| 14 | 0 | 2 | 0 | 0 | 0 | 0 | 0 | 0 | 0 | 2 | 2 | 0 | 0 | 0 | 0 | 2 | 13 | 0 | 0 | 0 | 2 | 0 | 0 |
| 15 | 0 | 2 | 0 | 4 | 0 | 0 | 0 | 2 | 0 | 2 | 2 | 2 | 2 | 0 | 0 | 2 | 9 | 0 | 0 | 0 | 0 | 0 | 0 |
| 16 | 0 | 0 | 0 | 0 | 0 | 0 | 0 | 0 | 4 | 0 | 0 | 0 | 11 | 0 | 0 | 0 | 0 | 0 | 0 | 2 | 0 | 4 | 0 |
| 17 | 0 | 0 | 0 | 0 | 0 | 12 | 0 | 0 | 2 | 0 | 0 | 0 | 2 | 0 | 0 | 0 | 5 | 2 | 5 | 0 | 7 | 0 | 0 |
| 18 | 0 | 0 | 0 | 0 | 12 | 0 | 0 | 0 | 9 | 0 | 0 | 0 | 3 | 0 | 0 | 0 | 0 | 0 | 29 | 0 | 0 | 0 | 0 |
| 19 | 0 | 0 | 0 | 0 | 0 | 24 | 0 | 0 | 0 | 0 | 0 | 0 | 6 | 0 | 0 | 0 | 0 | 30 | 0 | 0 | 6 | 0 | 0 |
| 20 | 0 | 0 | 0 | 0 | 3 | 12 | 0 | 0 | 0 | 0 | 0 | 0 | 0 | 0 | 0 | 0 | 3 | 15 | 9 | 0 | 27 | 0 | 0 |
| 21 | 0 | 16 | 3 | 0 | 0 | 0 | 0 | 0 | 0 | 0 | 0 | 42 | 3 | 3 | 0 | 0 | 0 | 0 | 0 | 0 | 0 | 0 | 0 |

Fig. 5. Profiles of individual TOM-clusters (rows) in terms of the VSM-clusters (columns). Cells indicate % of row totals.

We can illustrate this tendency by exploring the content of a particular set of clusters that are clearly related, as being wittnessed by the cross-tabulation of the two clusterings shown in Fig. 5, throughout the TOM- and VSM-based case. The TOM-based topic shown in Fig. 6 is characterized by the theme of modelling and explaining the behavior of ecosystems (TOM-cluster 5). On Fig. 7, the two VSM-clusters corresponding to this TOM-cluster are depicted (VSM-clusters 6 and 12). The topics connected in the previous case under the theme of ecological modelling are clearly recognizable in the two clusters, but being sharply separated into different and, in this sense, unrelated research directions, one focusing on the study concerning the role of species in ecosystems (*keystone species problem*, *foodwebs* etc.) and the other addressing the research on the mathematical modelling of such roles (*Lotka-Volterra models, equations, asymptotic stability* etc.). That is, the TOM-methodology seemed to be able to recognize the linkeages between narrower topics, and combine document sets so that these linkeages became visible. In sum, the results show that, in comparison with standard document clusterings, the overlay-map-based method leads to a much more informed grouping of papers into research lines, mirroring the dominant interrelations of basic topics.

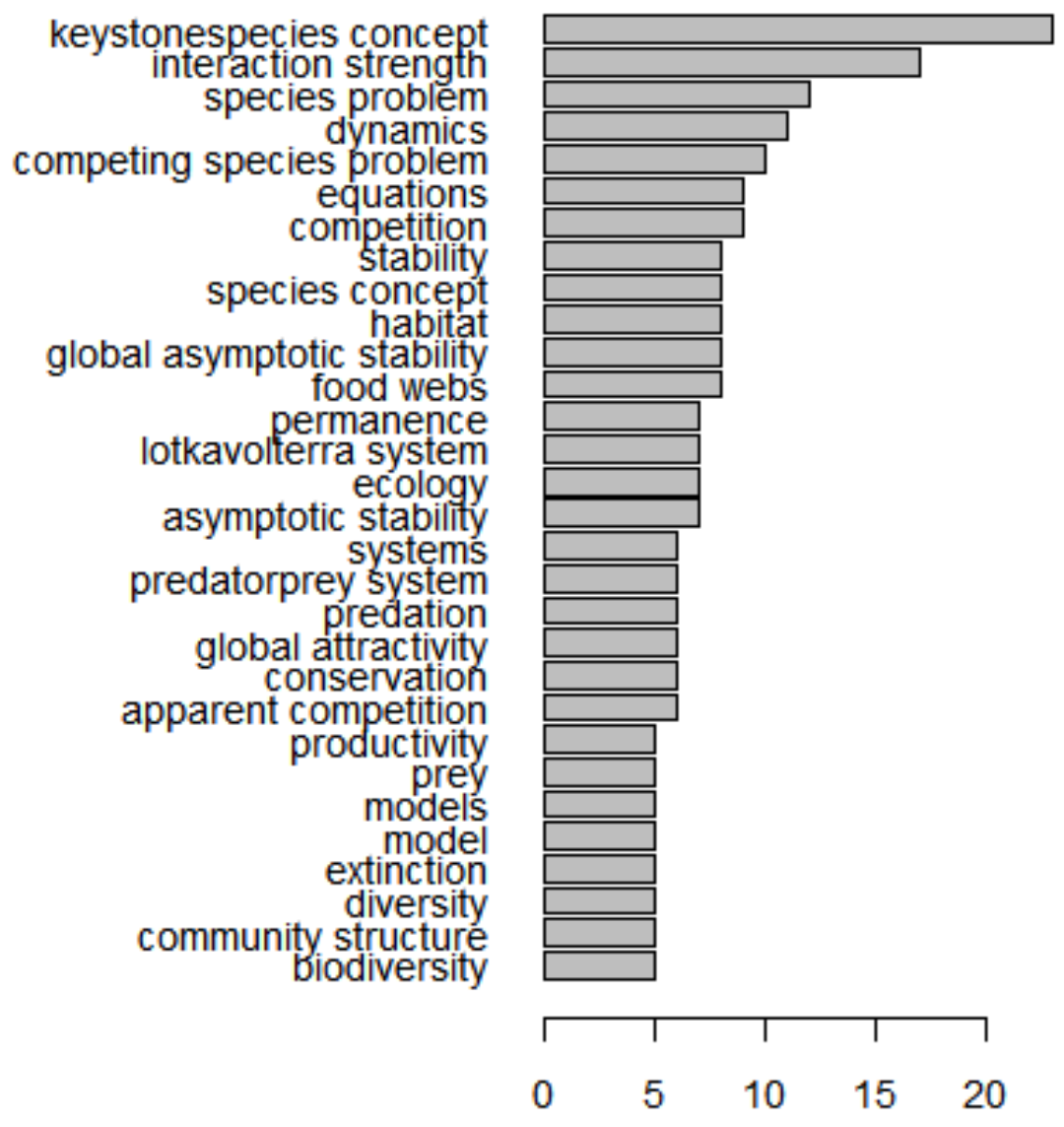

Fig. 6. The "ecosystems"-related cluster from the TOM-clustering

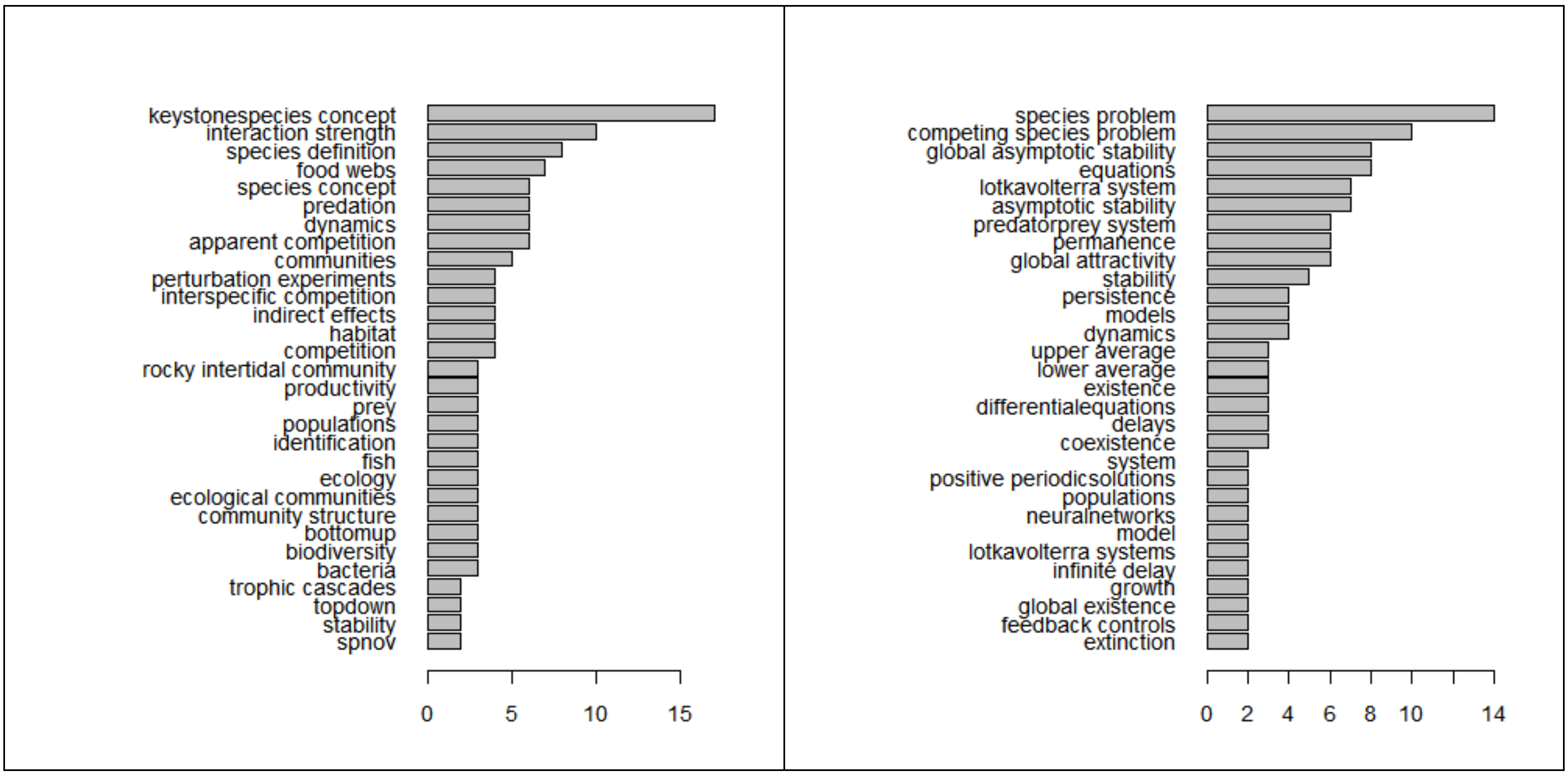

Fig. 7. Clusters overlapping with to the "ecosystems"-related TOM-cluster based on the VSM-clustering

*The topical structure and interdisciplinarity of the Species Problem*

By fully exploiting the analytic capabilities of the TOM framework we can gain a deep insight into the organization of the Species Problem, including the history of the discourse. To that end, we have devised a TOM-based profile for each topic cluster resulting from the clustering exercise above, composed of the following analytics:

- *The topic overlay map of the cluster*. As the central feature of the proposed methodology, the topic overlay map (TOM) is shown for each cluster. TOMs are depicted via the topic map (basemap) of the whole discourse (Species Problem) with the size of topics (nodes) being proportional to their relative share within the cluster (customized basemap). Consequently, TOMs make the dominant topics and their network, that underlie particular clusters, visible, so that the cognitive basis of the cluster would become transparent. Furthermore, since overlay maps characterize each document cluster in the context of the whole discourse (on the basemap), the internal position and relations of the cluster within the cognitive organization of the Species Problem is also made visible.
- *The keyword profile of the cluster*. As for the comparisons between TOM- and VSM-clusters reported above, a content profile was also generated for document clusters in terms of their most frequent keywords (frequency distribution). The keyword profile is based on author keywords and keywords generated from the titles of document reference lists.
- *The development of relative cluster size along the timeline*. A highly informative view on the evolution of the problem can be obtained by tracking the development of each cluster along the timeline. From this perspective, the internal trends of constituent research directions, the emergence, "rises and falls" of subdiscourses (encoded in document clusters) can be followed and compared, through which a dynamic and historical picture of cluster structure may be provided. Accordingly, an analytics is provided for each cluster combining three time series: (1) the annual—relative—size of the cluster, that is, the annual % from the whole set of documents comprising the cluster, (2) an 5-year moving average of annual cluster size and, for the purposes of comparison (3) an 5-year moving average on the annual relative size of the *whole* corpus, as the % of docs for each year. In verbal terms, (1)–(2) conveys how the cluster unfolds within the history of the problem, while (3) shows how the whole problem is progressing in parallel. In this way, by comparing the progress of the whole discourse with that of the particular research direction within, the relevance of sub-discourses can be traced and historically located for the SP.

Cluster profiles built in this fashion are collected in the Appendix of the paper. Striking even from the first and quick overview of these analytics is a general feature of the cognitive–topical organization of the Species Problem, that is clearly represented by the profiles. Most document clusters are based on a combination of the same "central" topic, depicted by node no. 1, and one or two related topic(s), that seem rather specific to the cluster in question. Topic 1 is made up of the issues and concepts most central to the Species Problem in general, while each cluster-specific topic, as we shall see below, covers a well-recognizable context for this "core" (i.e. Topic 1). In other words, all clusters appear to share the core topic (which contributes to all at a variable rate), but are still distinctively characterized by a different perspective (field, theoretical context, context of application etc.). In the following, we organize the overview of profiles according this structural feature of the discourse, that is, by building a typology of the contexts for Topic 1, as evidenced by the overlay maps for individual clusters.

*Species problem(s) related to Ecology.* The most clearly distinguishable context (or, rather, family of contexts) of the Species Problem is outlined by Clusters 5, 6, and 13. Cluster 5 is dominated by a topic from the study and (mathematical) modelling of ecosystems (topic #9), whereby research focuses on the role of species (as building blocks) in the functioning of such systems (*keystone species concept, equations, dynamics, food webs, Lotka–Volterra system* etc.). Clusters 6 and 13 are marked by topic #11, which represents a related, but conceptually different ecological/environmental direction, *viz.* conservation biology. In this context, the species problem is interpreted as establishing measures of biodiversity (cf. key terms as *biodiversity, richness, umbrella species*), which largely depends on the recognition (and individuation) of species taxa. Therefore, these two “conservation”-clusters are more central to the Species Problem, as introduced here, than the previous “ecosystems”-cluster, which fact is also indicated by the respective overlay maps in two respects. On one hand, the “conservation”-clusters show a dual dominance—interaction—of both the core topic (Topic 1) and the specific context (Topic 11), while the “ecosystems”-cluster is characterized by the specific context alone (Topic 9), with the core topic—the central concepts and issues of the Species Problem—being much more suppressed. On the other hand, the topology of the topic map (basemap) is also indicative in itself, since the “conservation”-topic #5 is part of a densely related topic-group around the core topic #1, while the “ecosystems”-topic is much less connected to this group, being attached to the topic map only through the “conservation”-topic. The relation between the latter two, however, is strong: it can be said that the “ecosystems”-topic is related to the core problem with the mediation of the “conservation”-topic. A further sign of the differing relevance of the ecology-based clusters is exhibited by the timeline-diagramms. Compared to the overall development of the Species Problem (indicated by the red curve with data points), the trendline for Cluster 5 is following a different course, showing an earlier peak (end of the 90’s) and a moderate decline afterwards. On the contrary, both Cluster 6 and 13 follows the main trend more closely, the curves progressing along the overall, ascending trendline. This altogether shows that the “conservation”-clusters are being much more integrated with the main discourse on the Species Problem, while the “ecosystems”-cluster seems to be a separate but overlapping direction. To put it another way, one can detect the combination of different “species problems” here, that are still separable via the TOM-profiles.

*The Species Category and the ontology of species.* As to the nature of the Species Problem, the most representative clusters, namely Cluster 8, 14 and 15, are characterized by the interplay between the core topic and Topic #5. This latter topic is a collection of issues and concepts related to the ontological status of species taxa (*natural kind, individuals, essentialism, Darwin*) that of the species category (*pluralism*, *realism*), and, quite tellingly, terms related to the school of systematics called “cladistics” *(cladist, cladogram, german, Hennig)*. The cluster profiles clearly show how the ontological—philosophical—issues penetrate into the mainstream biology-based discourse on the definition of species. In cluster 8, an equal weight is given to the core topic and the “ontological” topic, showing how the cladistic approach to systematics and the species concepts gains support from the thesis that species are ontological individuals (*species-as-individuals* thesis, SAI: cf. term frequencies *systematics, individuality, individuals, cladistics, monophyly, names, german*). A similar behavior can be attributed to cluster 14, with more weight given to "biology" (Topic #1). A further difference is that cluster 14 is expressed earlier in the timeline with a peak, preceeding (and anticipating) the rise of this subdiscourse (later also following an ascending timeline in itself). As opposed to these two, cluster 15 is marked by a clear dominance of the "ontology"-cluster over the core topic, which is also mirrored by its textual profile. This cluster coveys the theoretical (or meta-theoretical and philosophy-rooted) subdiscourse on the ontology of species and its implications on the species concept/category (*classification, natural kinds, individuals, evolution, pluralism, history* being the leading keywords).

The evidence for the proper interaction of philosophy and biology, wittnessed primarily by these profiles, also comes from both the overlay maps and the topology of the underlying topic map. The clear co-activity of the core topic and the "ontology"-topic is, though at a varying rate, universal for these profiles. Even more interestingly, the "ontology"-topic is not part of the dense topic group around the core in terms of network topology, just as the "ecosystems"-topic. Rather, ontological issues relate directly to the core, which links those issues to this central topic group. This picture also confirms the peculiarity of the situation, that an "outsider" discipline—philosophy—directly affects a scientific discourse, otherwise embedded in complex biological context. Furthermore, this arrangements can be taken as evidence not only for the interplay of distant topics, but, on top of that, for proper interdisciplinarity exhibited by the subject matter.

A further evidence for this deep embeddedness of the ontological perspective within the discourse can be drawn from the timeline diagramms. It can be seen for basically all three clusters that their progress goes "hand-in-hand" with the main trendline, reporting a shared dynamics of the general problem and the biophilosophical debate. The profiles alltogether well support the hypothesis behind the factors of the modern SP: in particular, it is made visible how the individuality thesis affects the success of the so-called cladistics-based (and, in derived forms, phylogenetic and genetic) definition of species.

*The species problem and specialities of bioscience.* Many profiles resulting from this mapping can be grouped into a well-defined family of clusters (or subdiscourses). The feature that collects these profiles together is that each cluster is concentrated around the realization of the species problem within a speciality of bioscience—the latter usually focusing on a particular, but broader taxonomic group (such as fungi or algae). A common characteristics for most of these specialities is that standard species concepts (in fact, the notion of species) is problematic for their purposes, due to the special nature of the subject matter (taxa of interest). Cluster 9, with Topic #12, mostly represents the problems micology in defining species out of fungi; Cluster 11 and Topic #14 stands for the species problem in botanics, plants often exhibiting "irregular" speciation behavior and patterns against the Biological Species Concept (*reproductive isolation, hybridization, speciation, polyploidy, pollination* etc.); Cluster 16 with Topic #14 places the SP in the context of paleontology, whereby the reconstruction of species from the fossil record makes it hard to apply modern definitions (such as the Biological SP) based on "observable" relations between existing taxa (reproductive isolation). A deeply interlinked subgroup of clusters is concerned with microbiological taxonomy: Topic #3 and Topic #2 underlies a couple of profiles related to the taxonomy of algae, more specifically, diatoms, whereby it is extremely problematic to empirically systematize biological diversity (Cluster 17, Cluster 20). Cluster 18 is specifically focuses on systematic bacteriology, bacterial phenomena escaping most approaches to defining species in this realm, represented by Topic #0.

This collection of cluster profiles also exhibits an important structural aspect of the Species Problem. As can be seen from the outlines above, each cluster is a combination of the core topic and a cluster-specific topic. According to the overlays, each of these cluster-specific nodes belongs to the dense central topic group directly surrounding the core (Topic #1), in terms of network topology. That is to say, these clusters provide the main context in which the SP exits and develops. This structural feature provides further evidence that the Species Problem has been, and is being mainly fed by the problems of applying theoretical concepts in the practice of biology ("field research"). To put it another way, this arrangements shed light on a main dimension of the SP, viz. the difficulties to achieve the theory-driven goal of formulating a species concept that is universal enough to cover the diversity of the living world. The timeline diagramms also support this interpretation by indicating a high fit between the main trendline and the respective cluster curves.

*The Species Problem and methods from molecular biology.*– Still part of both the speciality-driven clusters is Cluster 10 focusing on the delimitation of species taxa with the aim of molecular biology. The distinctive topic (Topic #2) unerlying this cluster is also part of the central group of basemap nodes. The reason for still treating this profile separately is twofold: instead of specific taxa, this subdiscourse is concerned with "methodological paradigm" that cross-fertilizes taxonomic schools and different approaches to the species concept, and, nevertheless, represents the present "instrumentalist" consensus on the species problem. Relying on genetic and molecular markers (cf. *RNA secondary structure, ribosomal RNA, mtDNA, molecular phylogeny, polymerase chain reaction*) to separate species by inferred phylogenies is a pragmatic approach that accomodates features from many theoretical species concepts (Phylogenetic, Cladistic, Genetic, Biological), while practically overlooking conceptual problems. This cluster can, therefore, be seen as the response of normal science to the theoretical debate with an "inference to the best explanation"-type framework.

*The Species Problem and evolutionary mechanisms.* Somewhat distributed or scatterred among clusters is a "horizontal" theme that can still be well recognized from browsing the profiles. This theme covers the research in evolutionary biology concerned with *speciation* and *speciation mechanisms*. This issue is inherently related (sometimes even hardly separable) to the definition of the species category. It is widely assumed that had the "natural" mechanism(s) that isolates species been found, a corresponding definition based on this/these mechanisms would also naturally follow. Cluster 12 heavily relies on this theme (*sympatric speciation, adaptive radiation, reproductive isolation, hybridization*), though with a focus on viruses with Topic #13 (belonging, in that sese, to the taxa-specific core group). It is also well-recognizable in Cluster 11 with Topic #14 (*reproductive isolation, hybridization*), which similarly furhter specialzes in the study of a specific taxon (plants).

## Conclusion

In this paper, a multi-purpose analytical framework, inspired by science mapping models, has been introduced, that aims at the in-depth exploration of scientific discourses. The Topic Overlay Mapping methodology has been shown to be capable of mapping scientific contexts in various dimensions: it provides a visual as well as a measureable map of the topical organization of the corpus, grounding visualization, trend analysis and document clustering. In other words, TOM aids the structuring of the corpus by uncovering the latent factors that account for its structure, simultaneously, enhancing and driving the user's abilities to explain the development of a large-scale body of literature.

An important motive of the approach comes from text mining and is related to application (2), that is, document clustering. A relatively new trend in document clustering/classification is based on the observation that the classical, so-called "Vector Space Model" (VSM) for grouping texts ignores the relationship of terms within the corpus, and also the fact that terms occur in overlapping or diversely related topics. In recent years, this aspect of cognitive complexity has been addressed by novel methodologies, such as graph-based classification or LDA (Latent Dirichlet Allocation) techniqes, both related to our framework. Graph-based classifications attempt to extract a relational structure of terms from the corpus to assist the grouping of documents (Wang et al. 2005), and LDA-applications assume overlapping topics (as term distributions) and texts as topic combinations. Unlike these graph-based methods, that usually belong to machine learning or "supervised" techniques, and whereby the graphs mostly convey grammatic or ontology-driven relations, our framework accomodates a non-supervised clustering based on the complex topical relations drawn from the corpus. On the other hand, as a multi-purpose framework, it also provides (via overlay maps) an intuitive visualization of topical relationships aiding the interpretation of the results, which is not the case for LDA or similar generative models.

Experimental results show that, in comparison with standard document clusterings, the overlay-map-based method leads to a much more informed grouping of papers into research lines, mirroring the dominant interrelations of basic topics. As a use case, we applied the TOM framework in a document clustering exercise to reveal the cognitive organization, as well as the historical development of a complex issue in biosystematics, the debate on the species concept. The procedure uncovered a series of important insights into this interdisciplinary discourse. The topic map (basemap), along with the document clusters based on its structure provided evidence that the conceptual problem has been mainly shaped by biological specialities for which mainstream species concepts (such as the Biological Species Concept) implied criteria hardly applicable in the corresponding biological domain. However, from the integration of topics within clusters, an equally important drive of the problem can be implied, which is the meta-level (philosophical) debate on the ontological status of both species and the species category. it is confirmed that, though the modern consensus is concerned with molecular methods to discover species, the roadmap to the state-of-the-art accomodates various philosophical “interventions”. The roadmap is reflected in the cluster profiles: the ontological claim for species taxa being individuals provided philosophical support for the phylogenetic and cladistic concepts of species, which, in turn, served as the theoretical basis for applying methods from molecular genetics to delineate species taxa (reconstructing molecular phylogenies). As such, the TOM framework confirmed that the history of the species problem is a case of strong interdisciplinarity, whereby distant disciplines interact to form new paradigms.

## Acknowledgment

This paper was supported the European Union and the European Social Fund through project FuturICT.hu (grant no.: TÁMOP-4.2.2.C-11/1/KONV-2012-0013); the European Union Seventh Framework Programme (FP7/2014-2017) under grant agreement n° 613202 (IMPACT-EV project) and by the János Bolyai Research Scholarship of the Hungarian Academy of Sciences.

APPENDIX

## Cluster 1

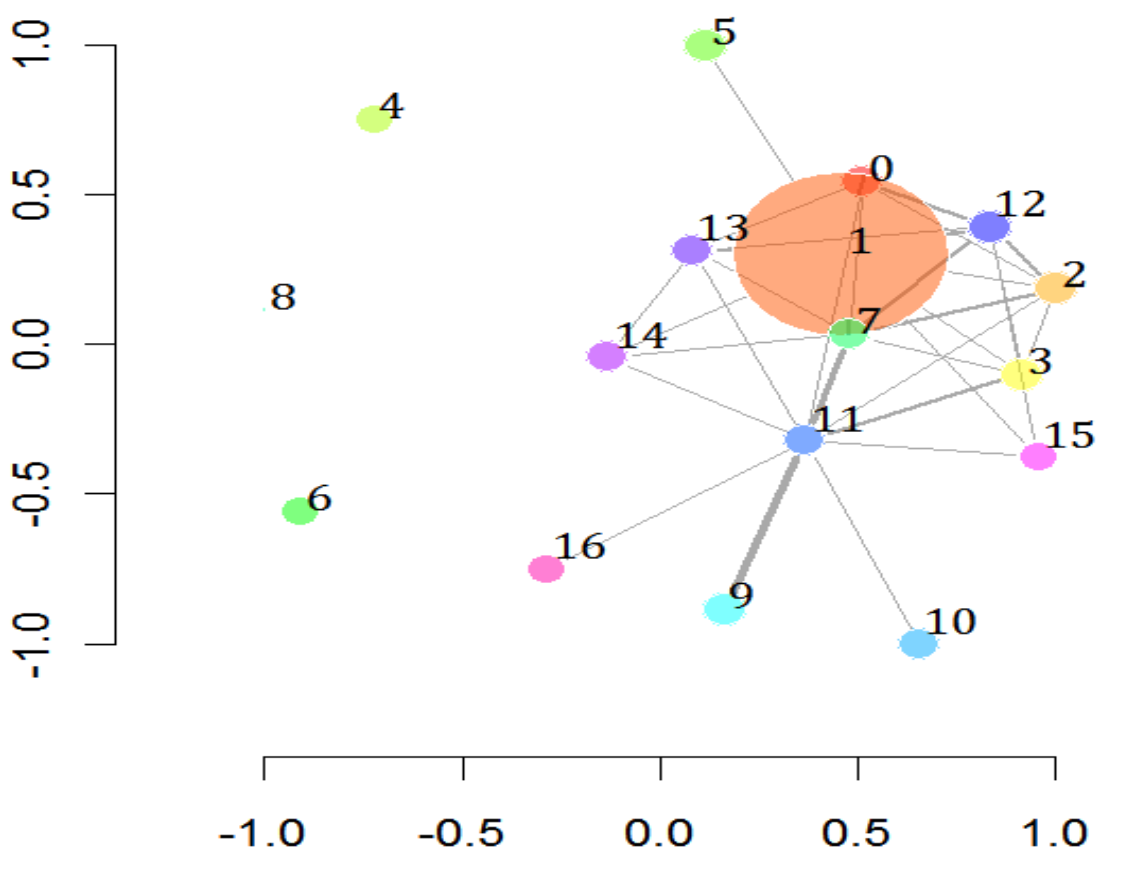

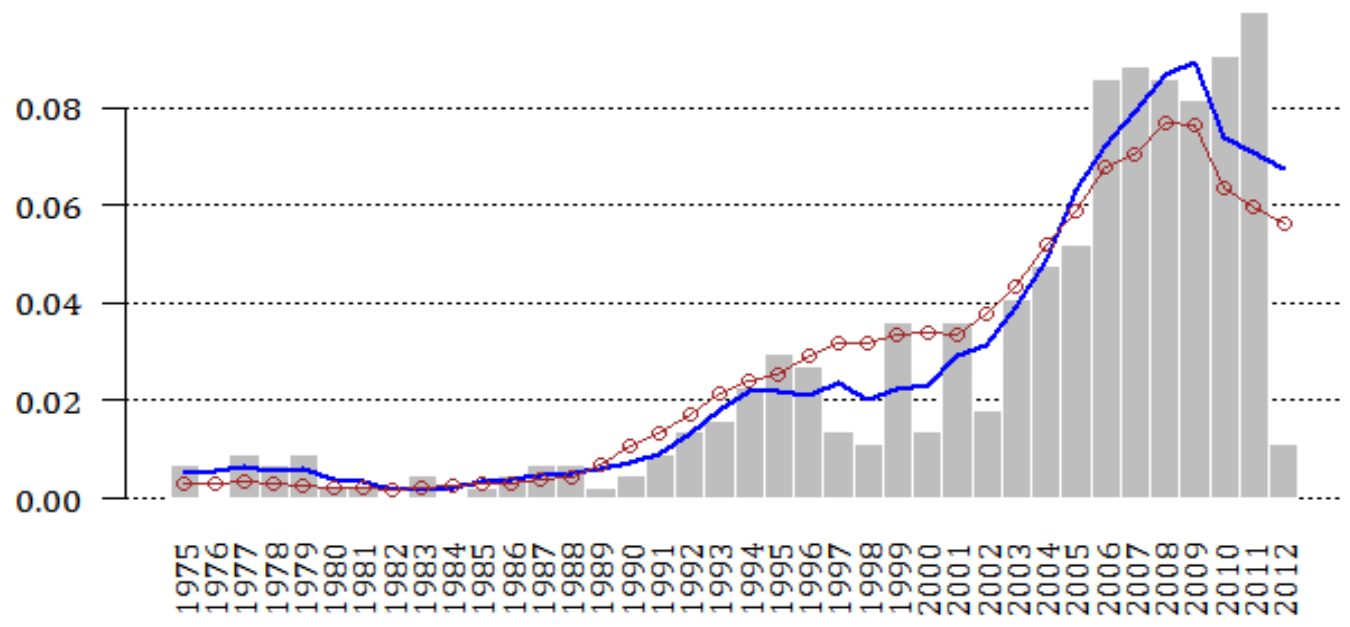

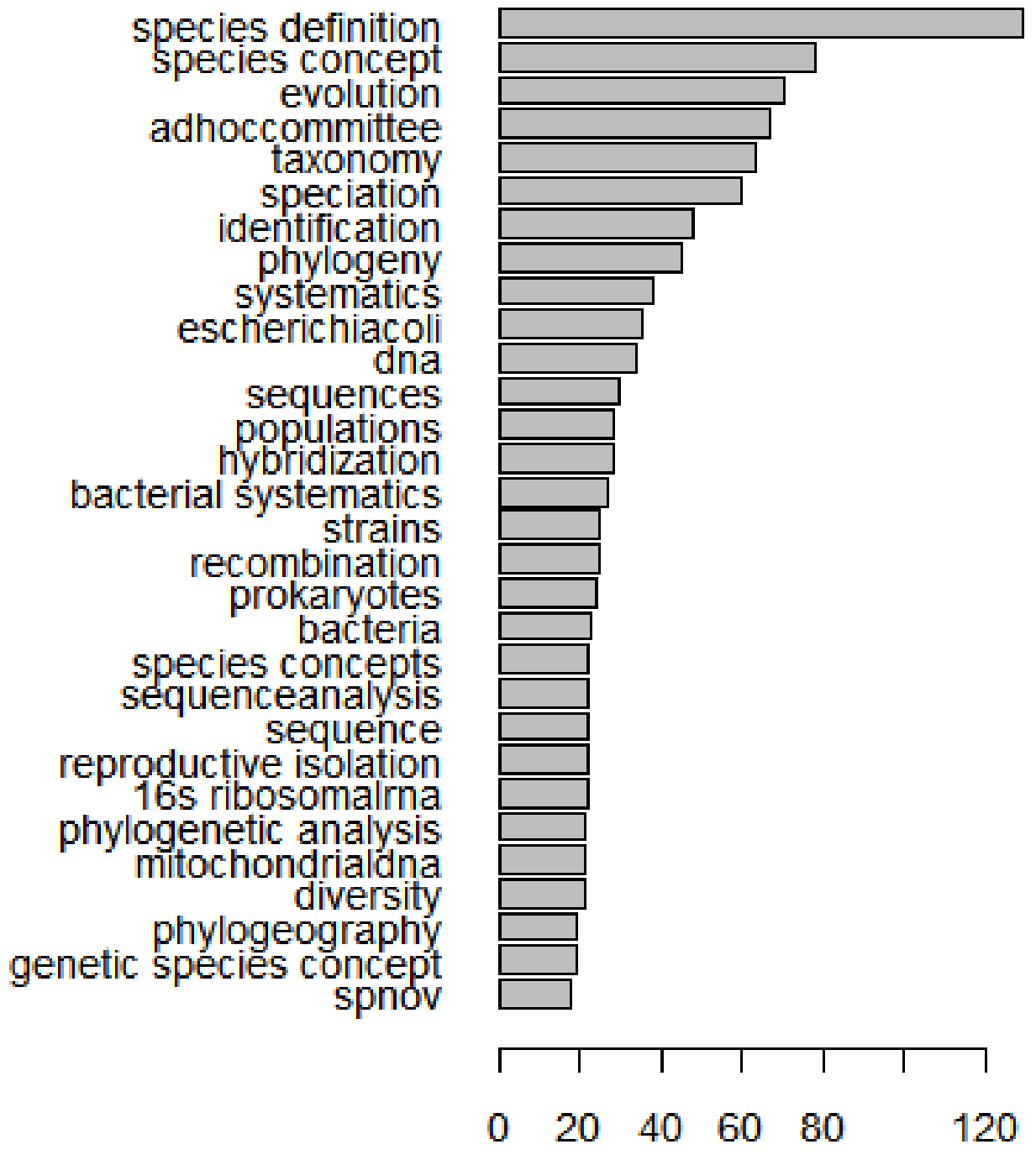

## Cluster 2

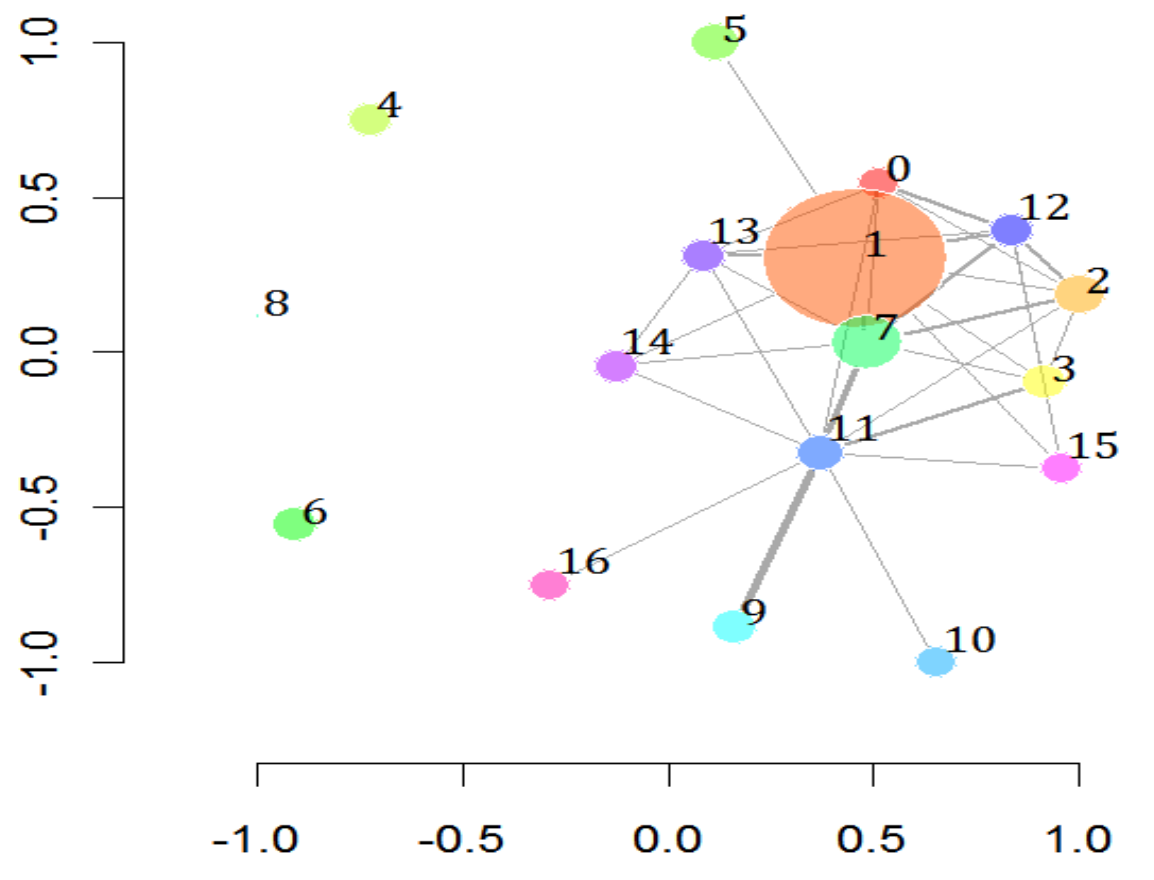

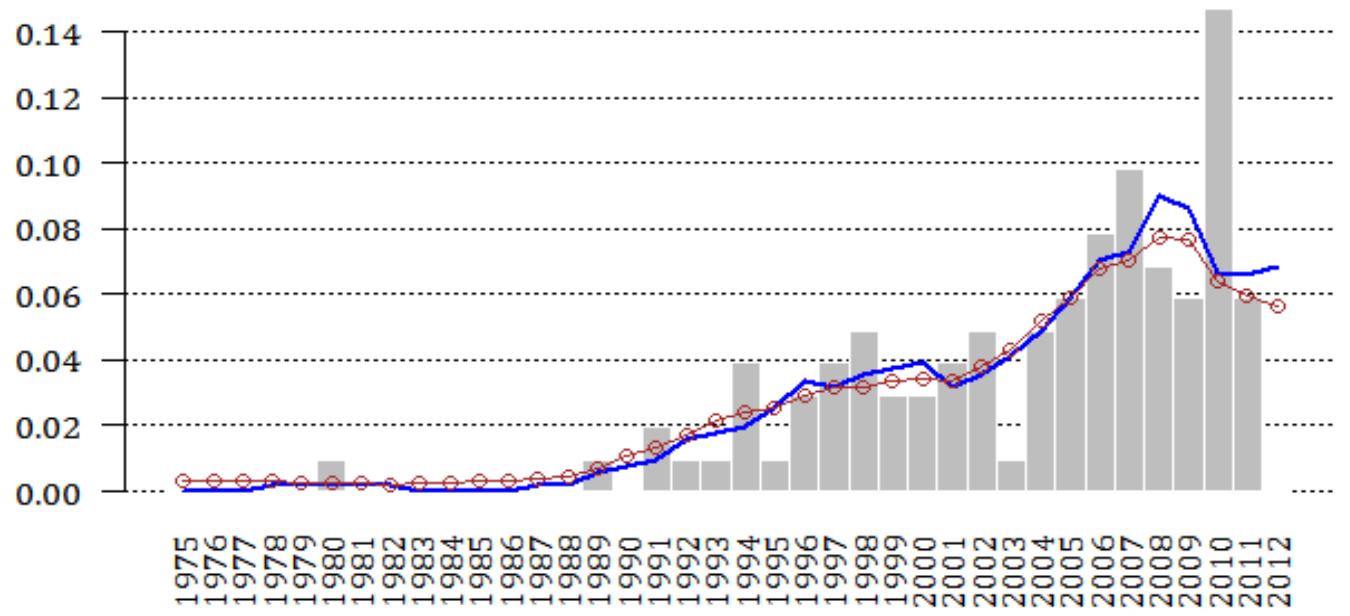

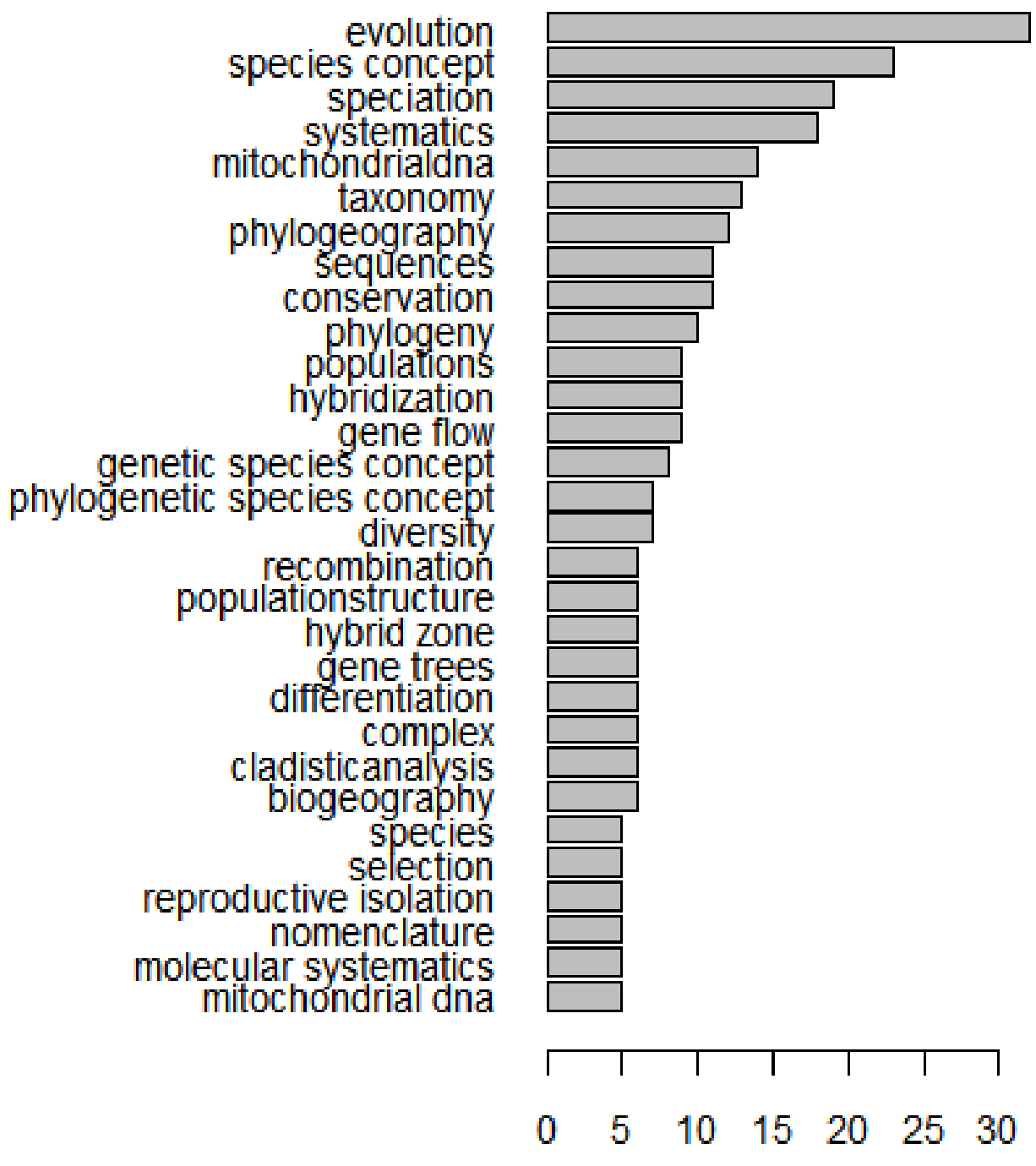

## Cluster 3

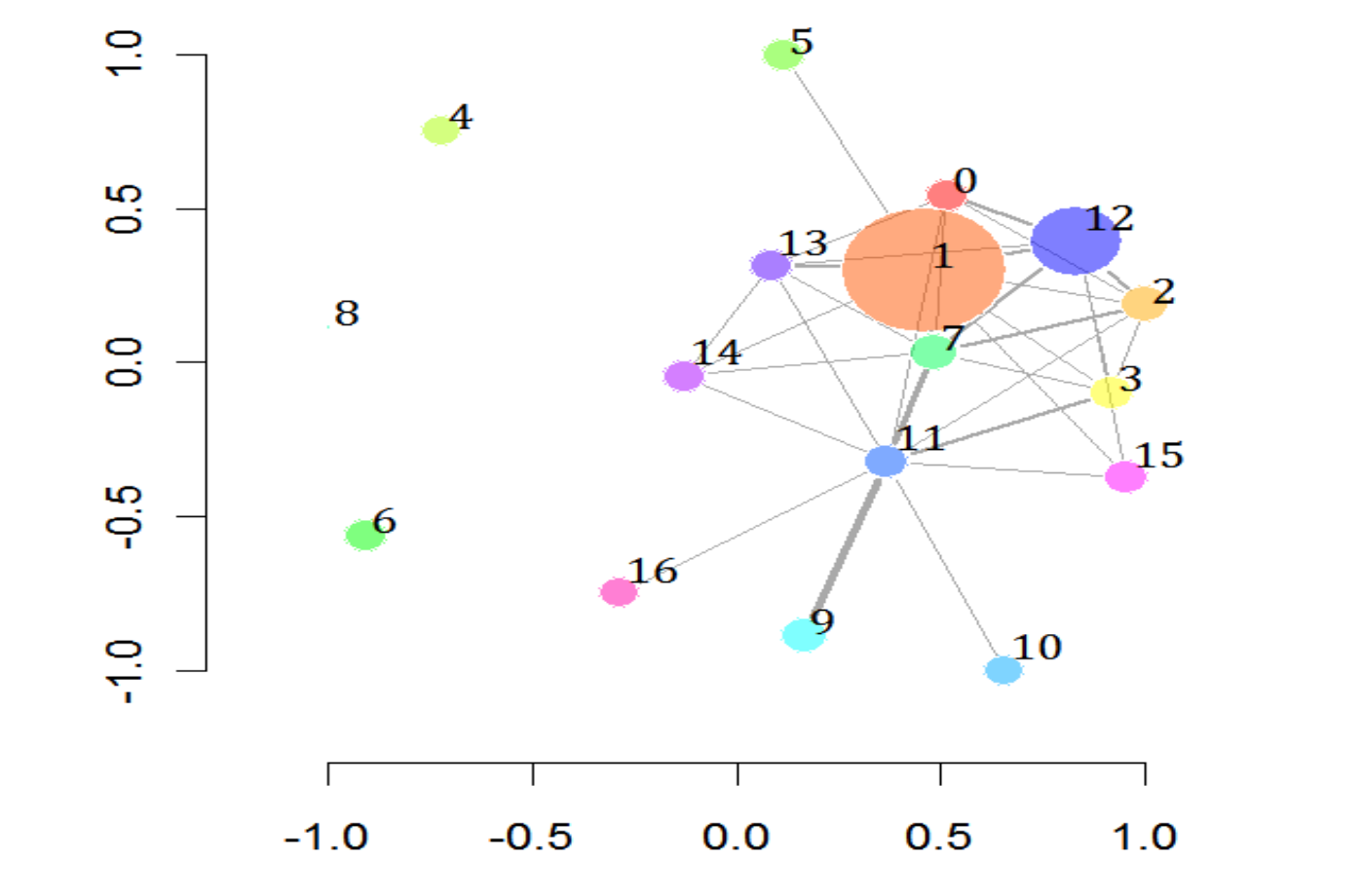

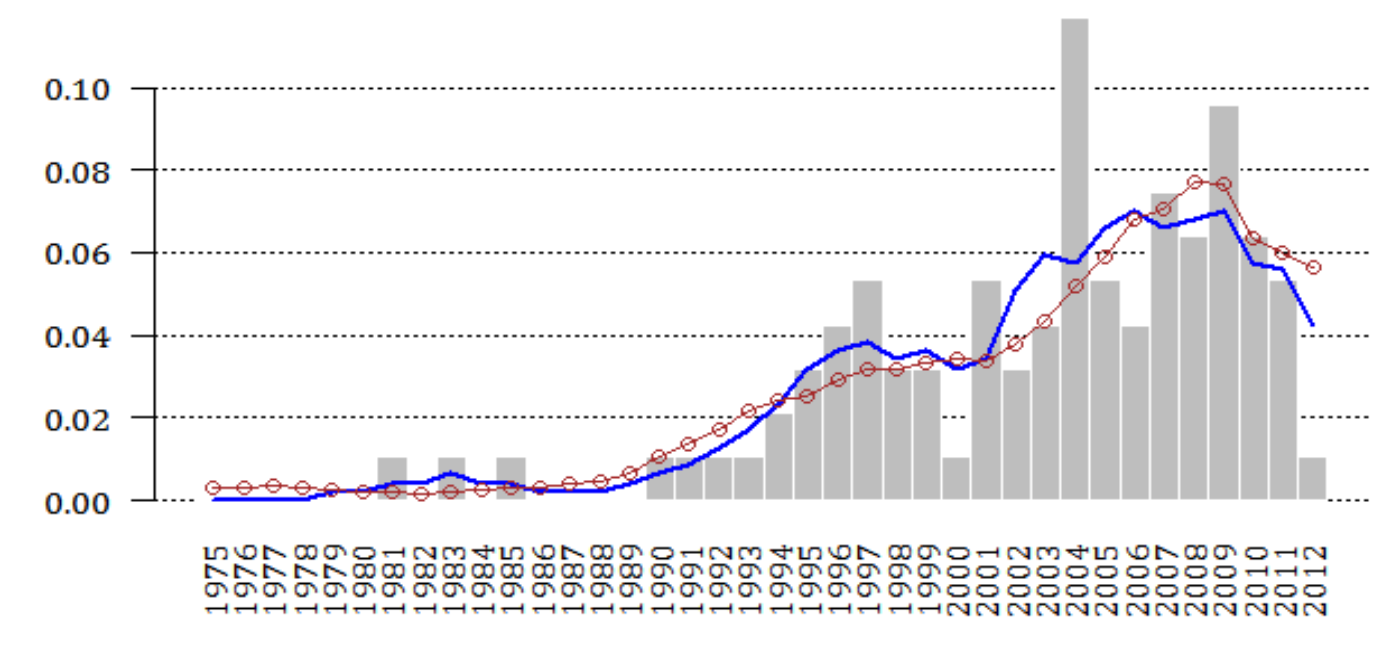

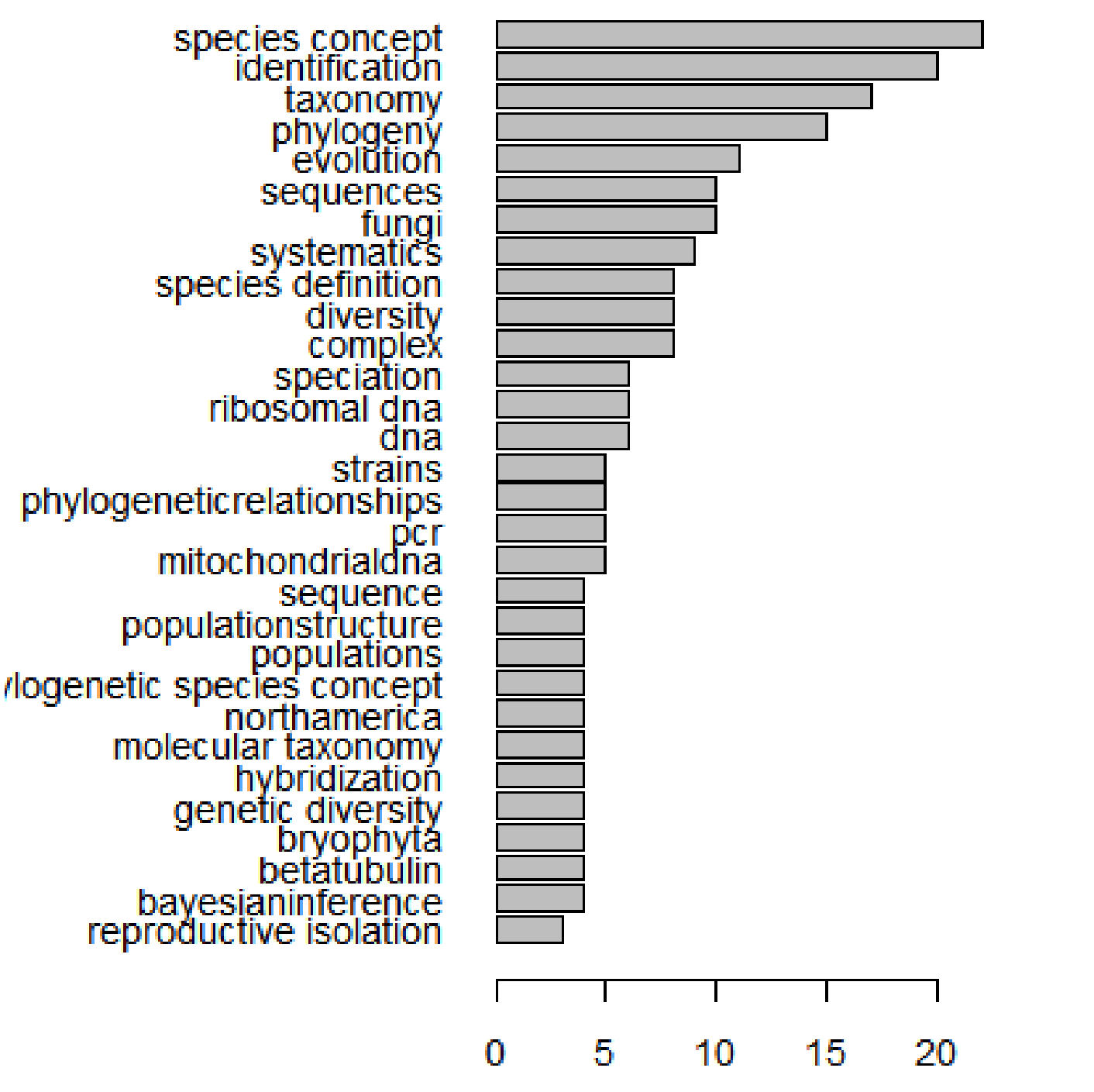

## Cluster 4

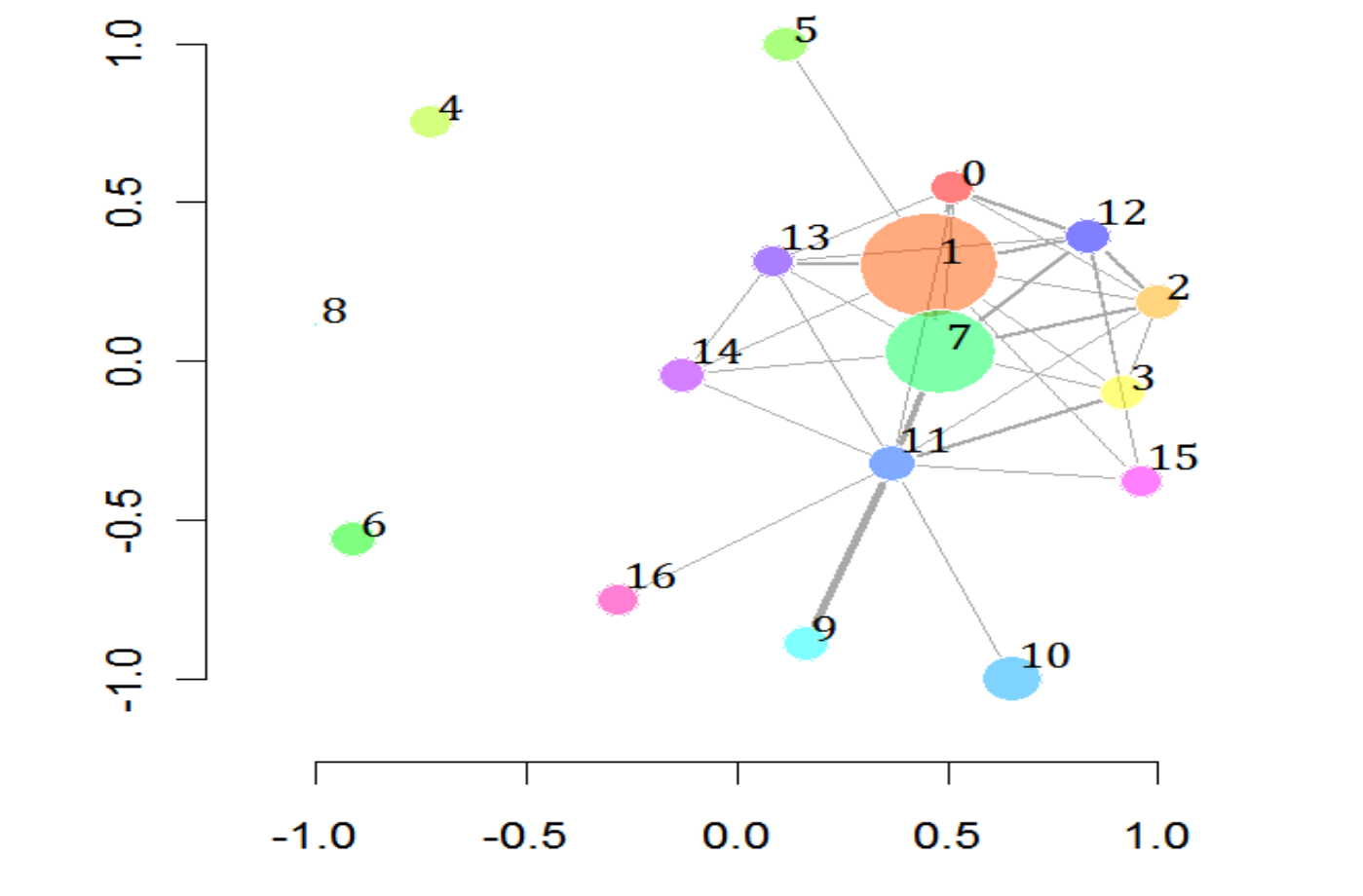

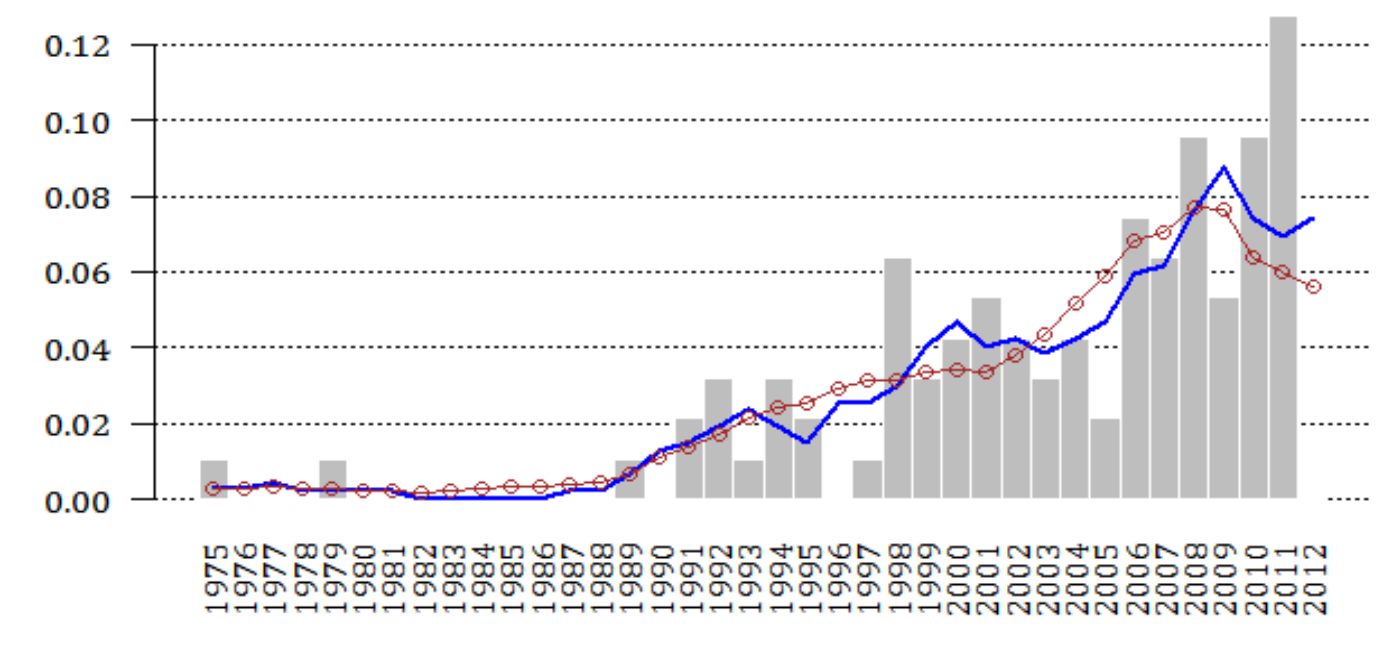

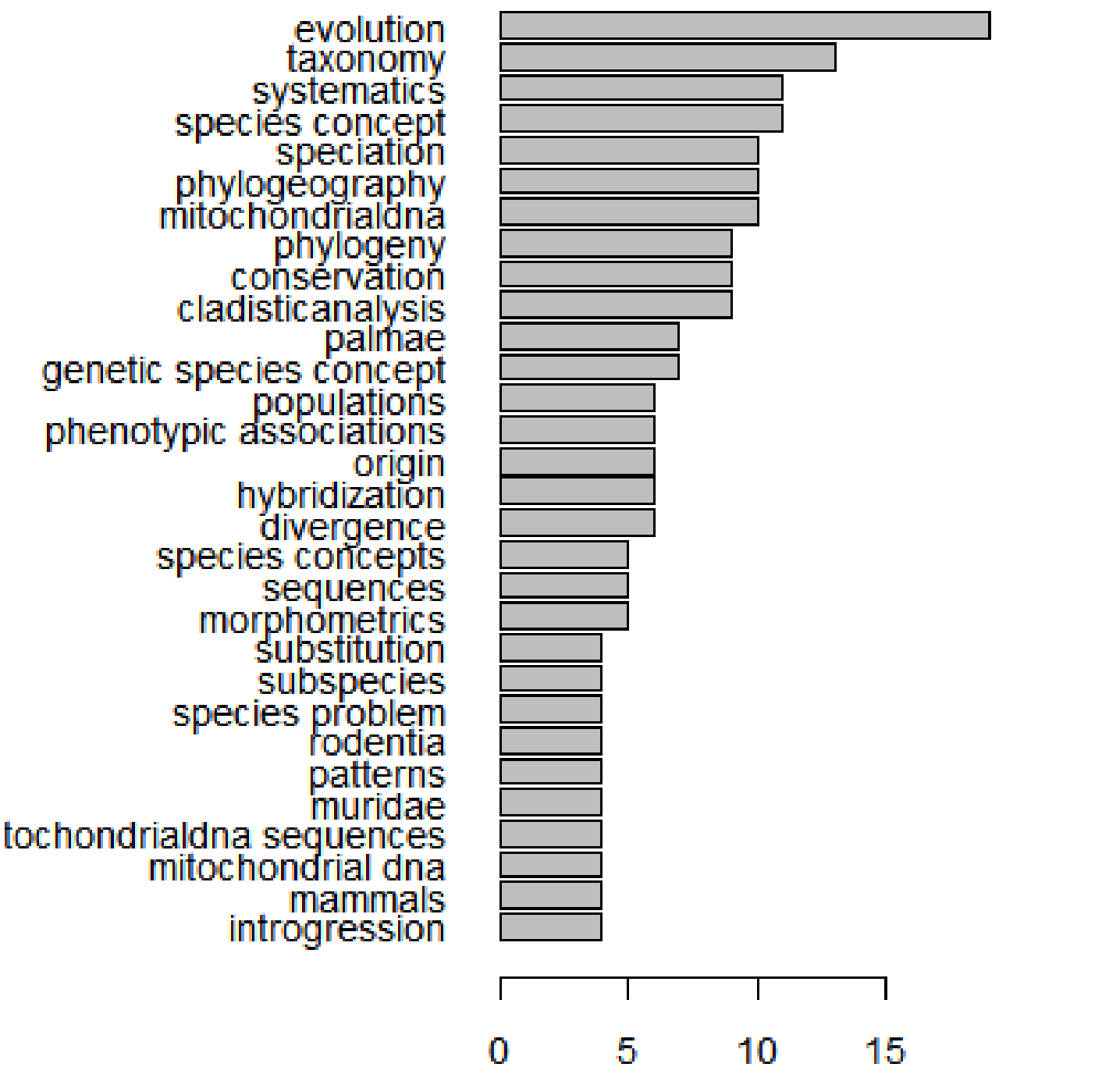

## Cluster 5

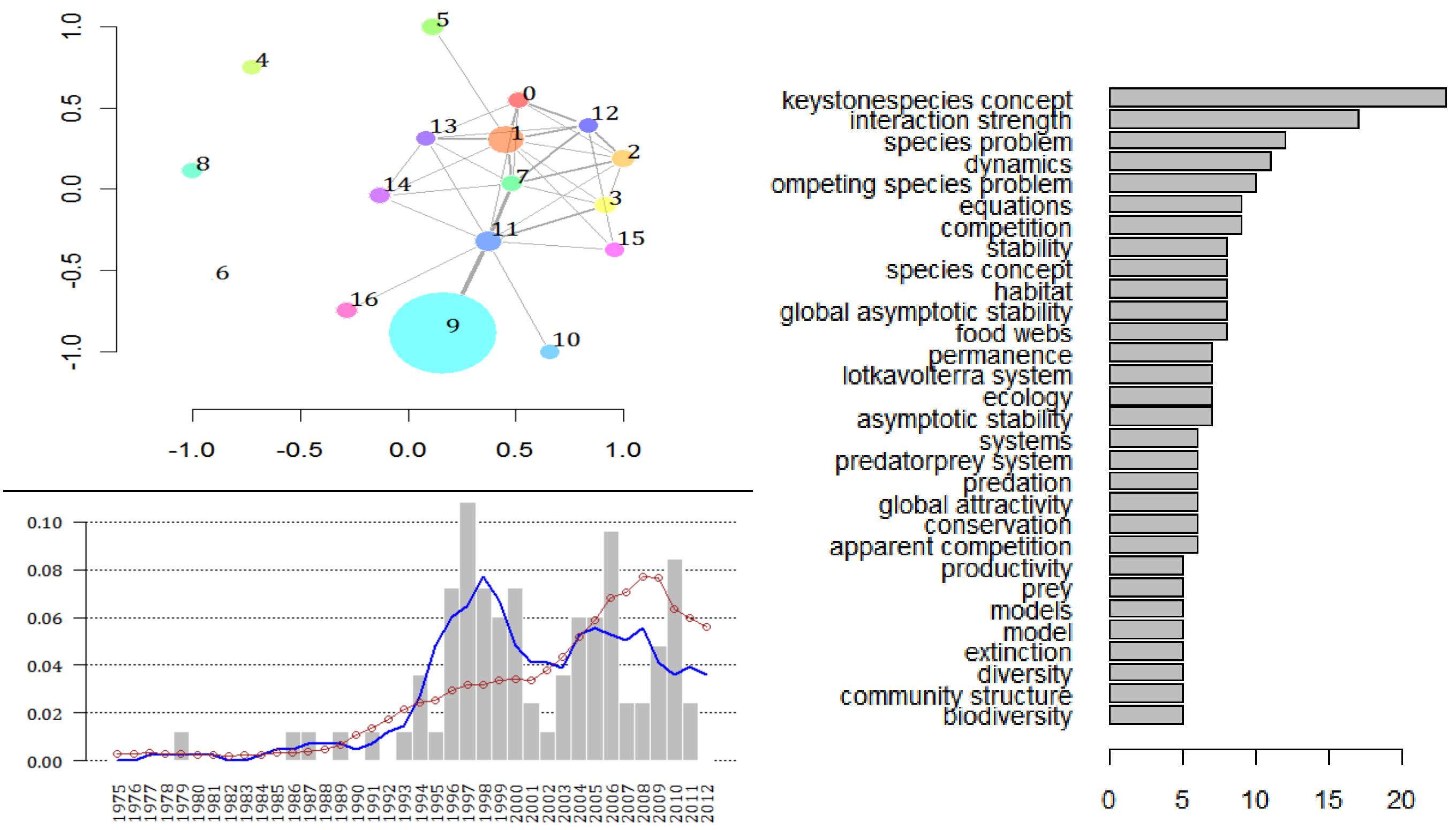

## Cluster 6

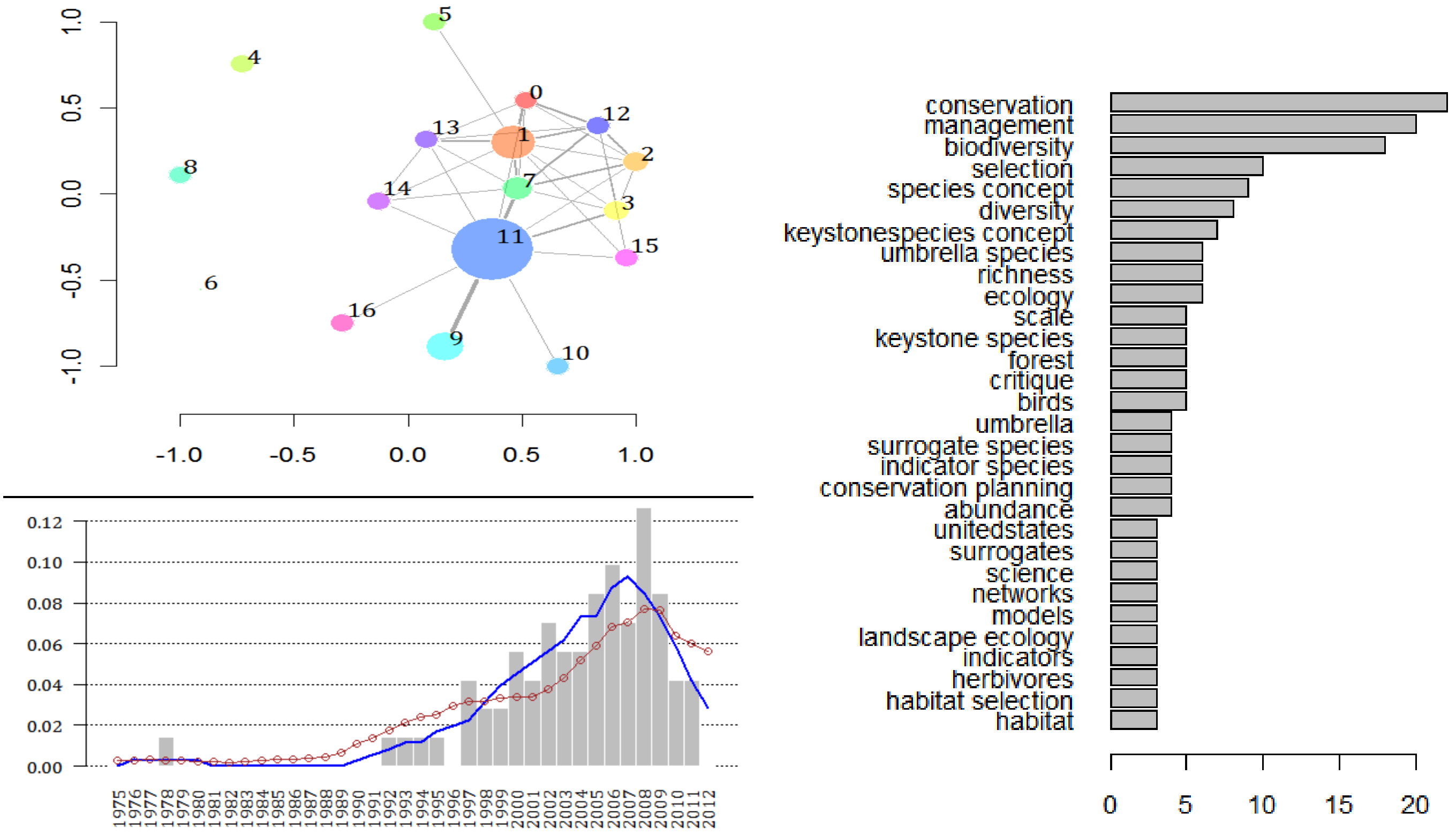

## Cluster 7

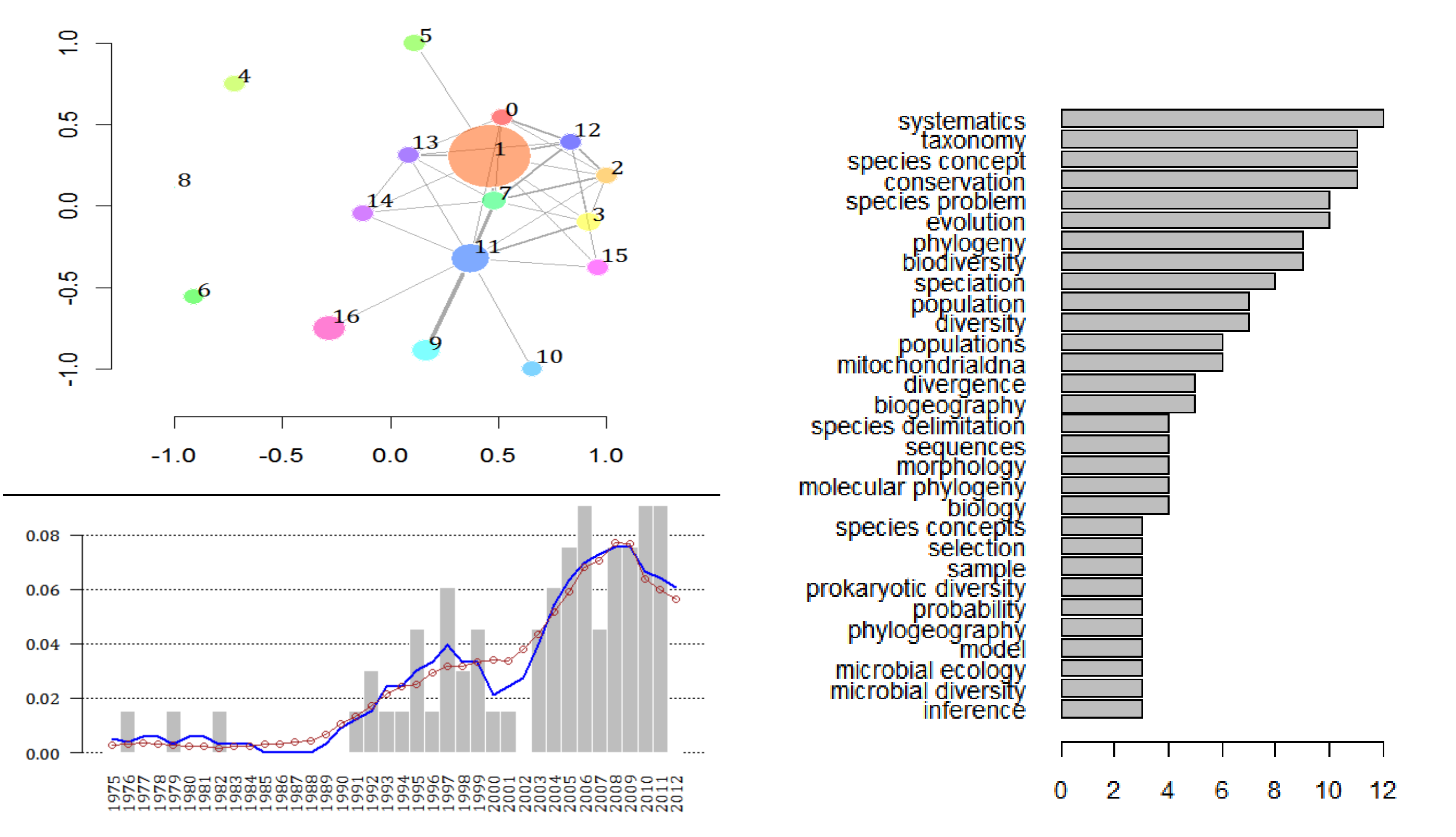

## Cluster 8

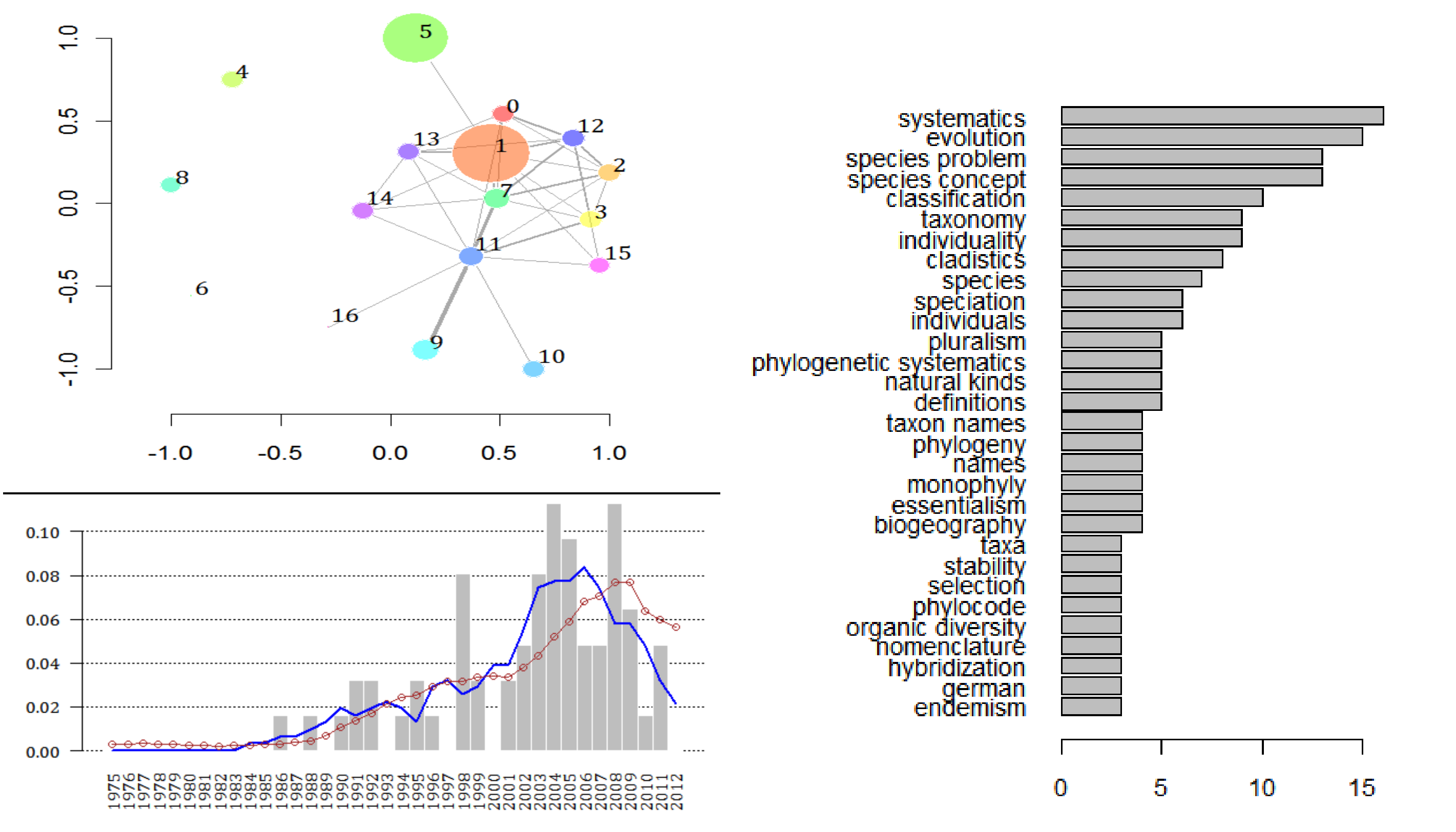

## Cluster 9

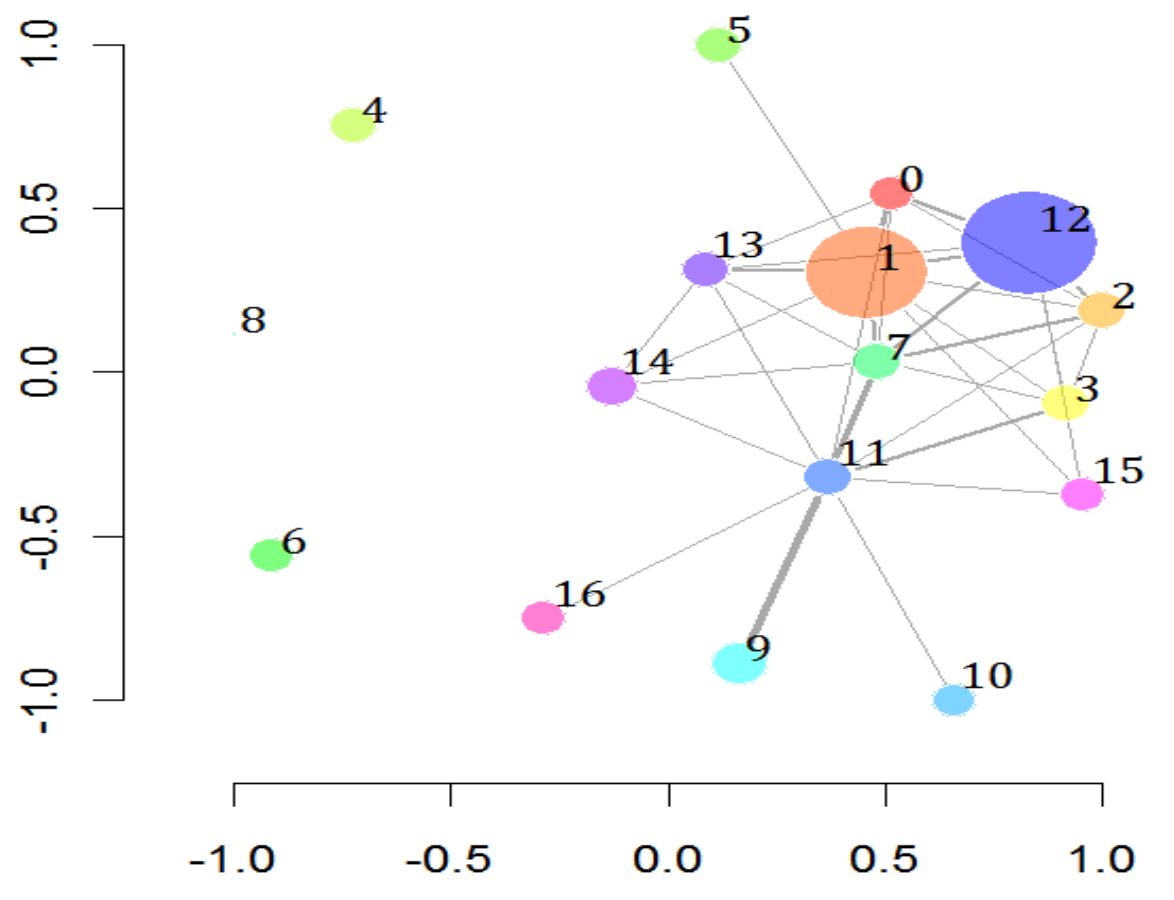

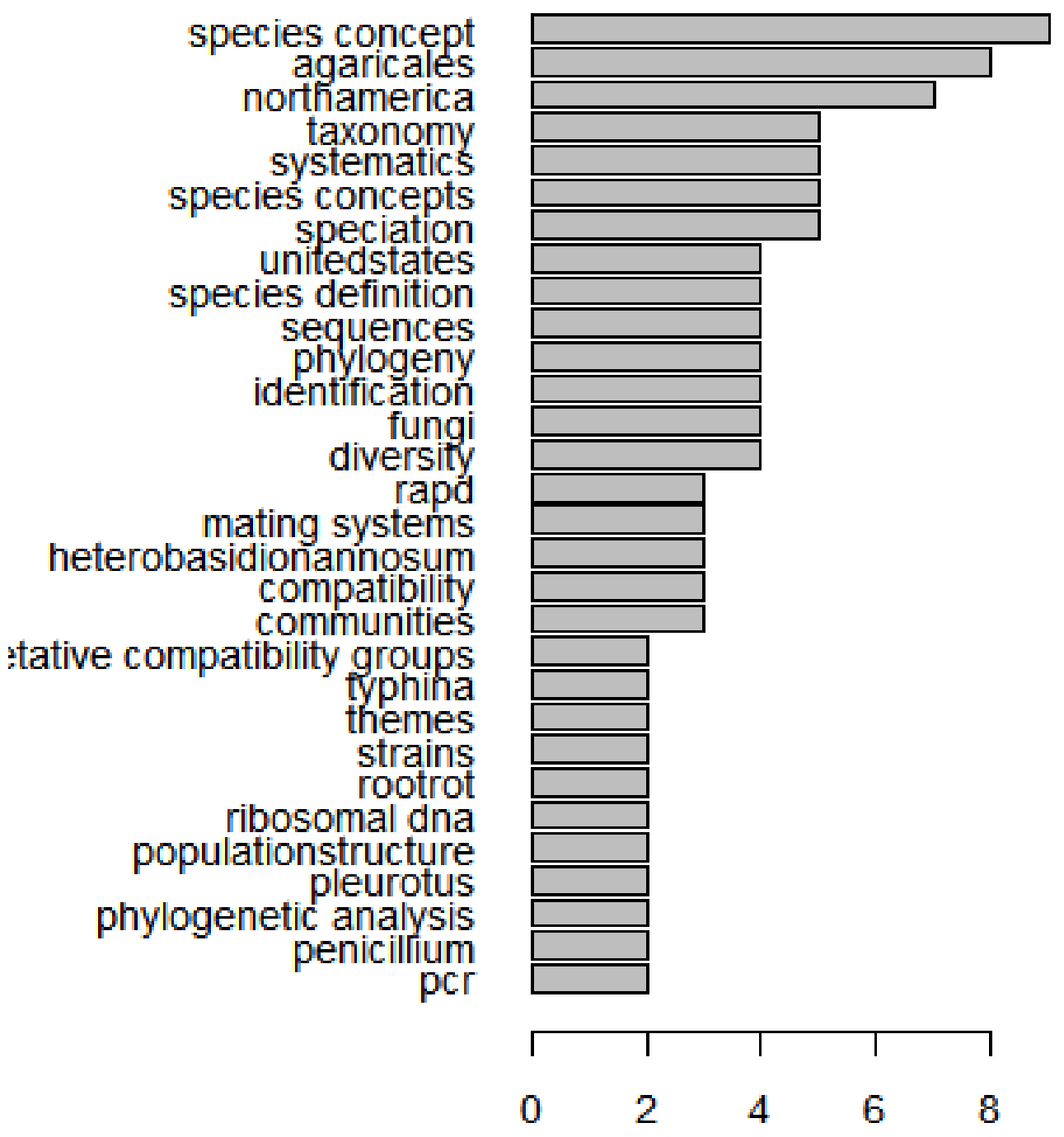

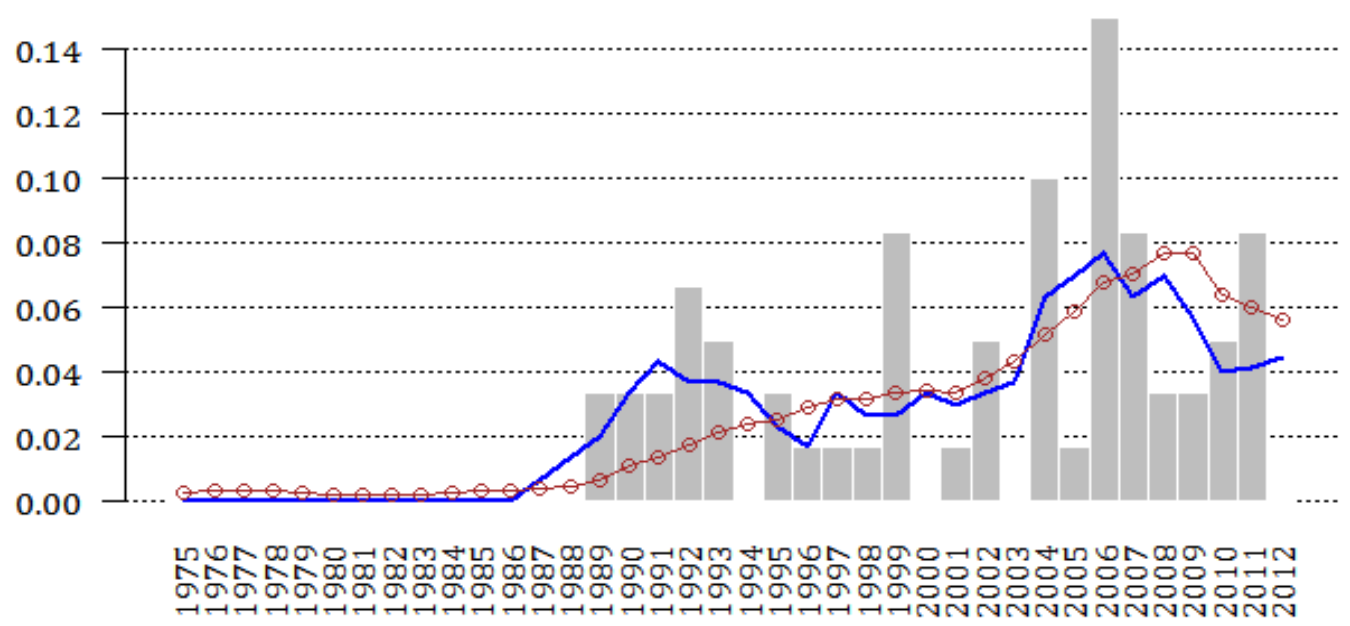

## Cluster 10

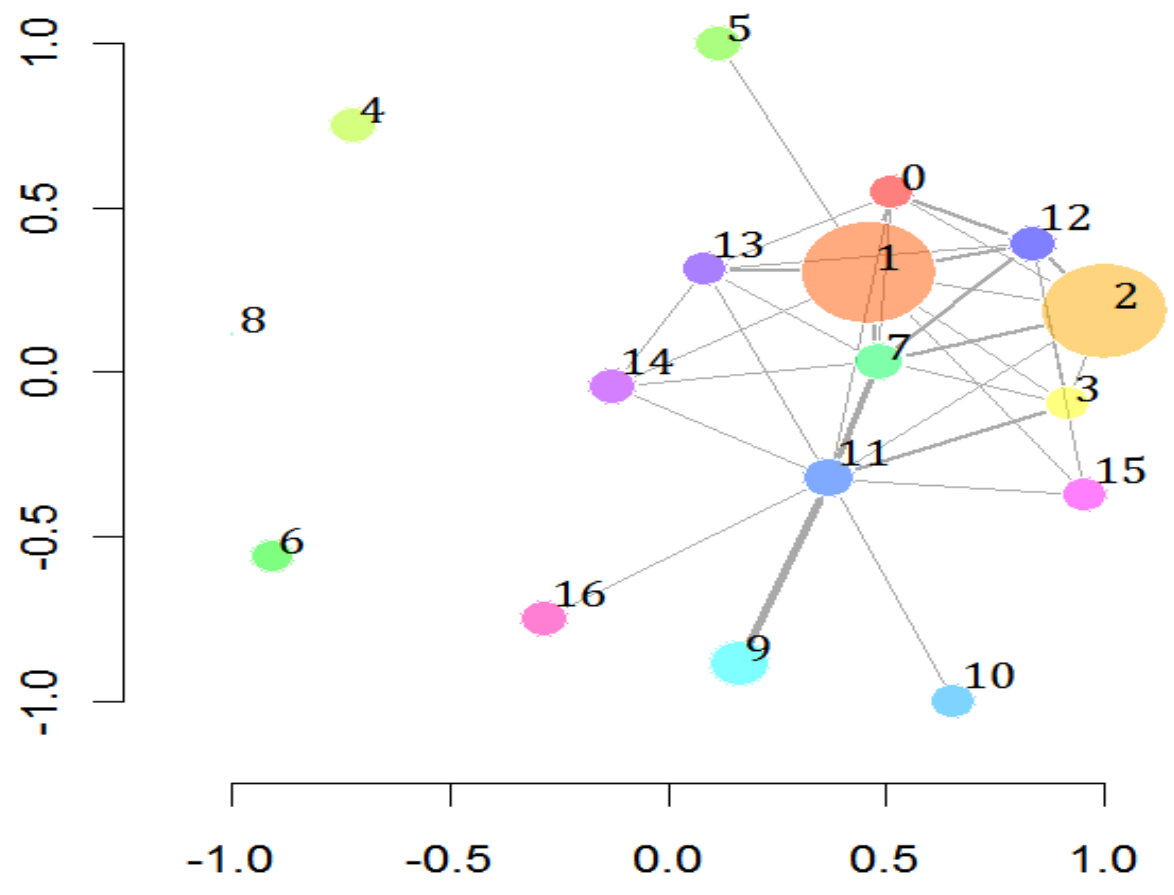

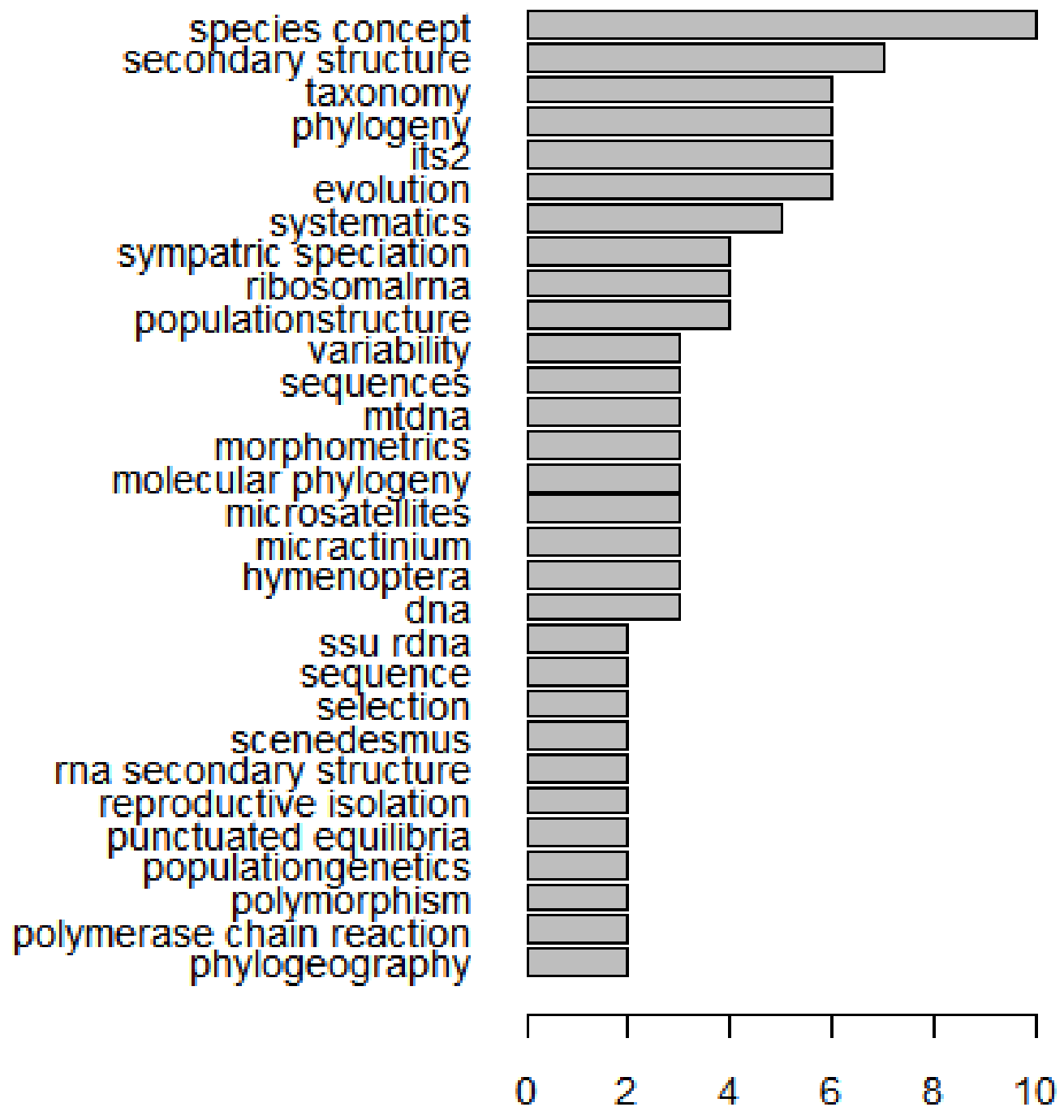

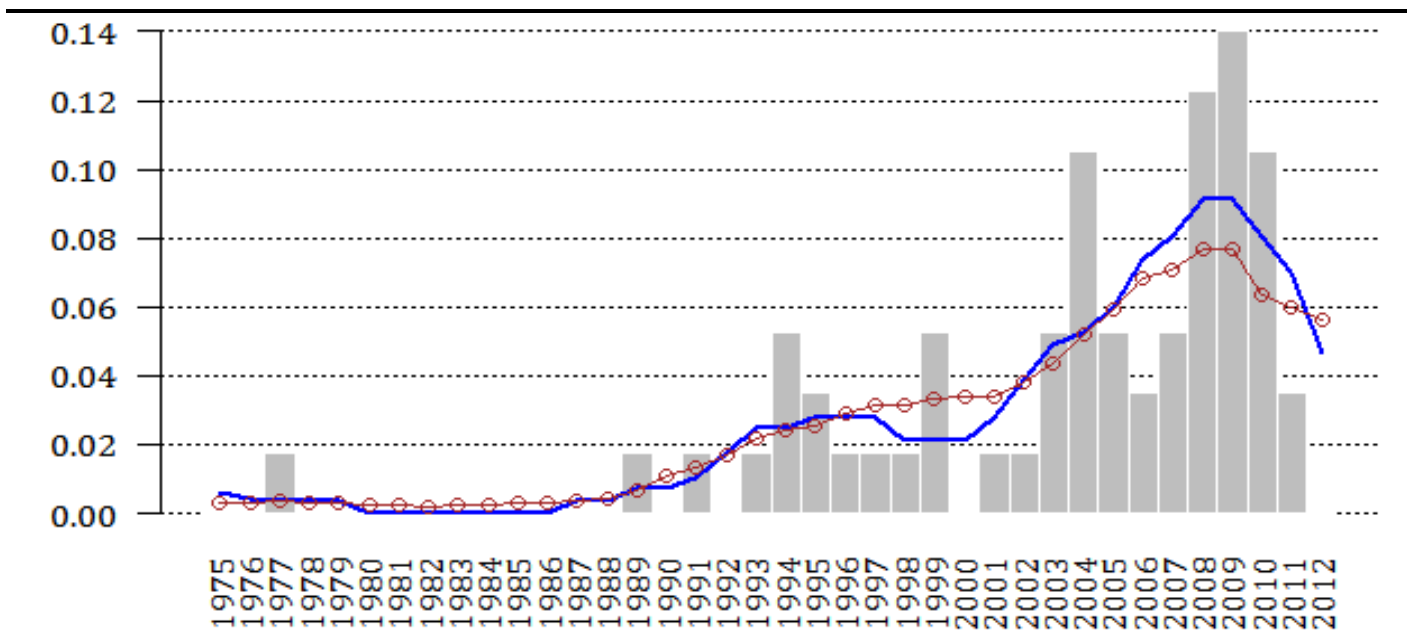

## Cluster 11

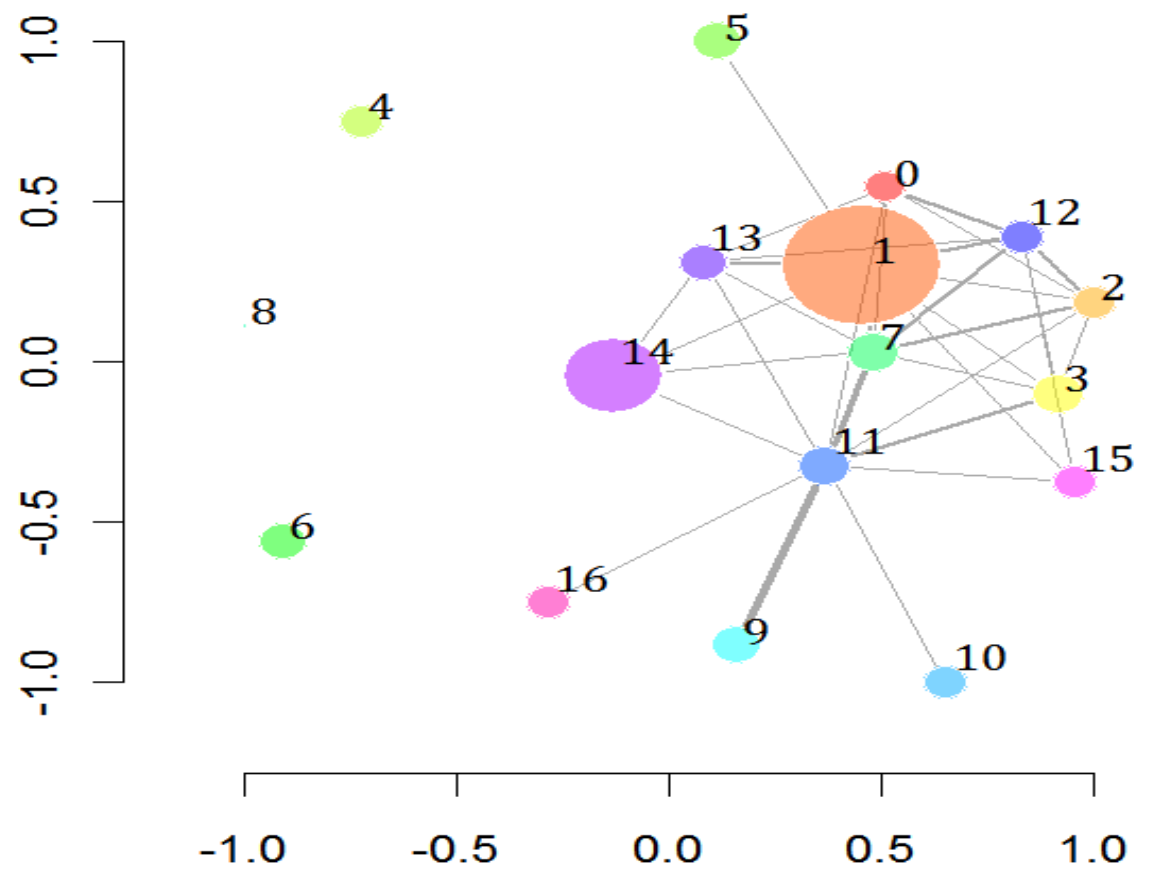

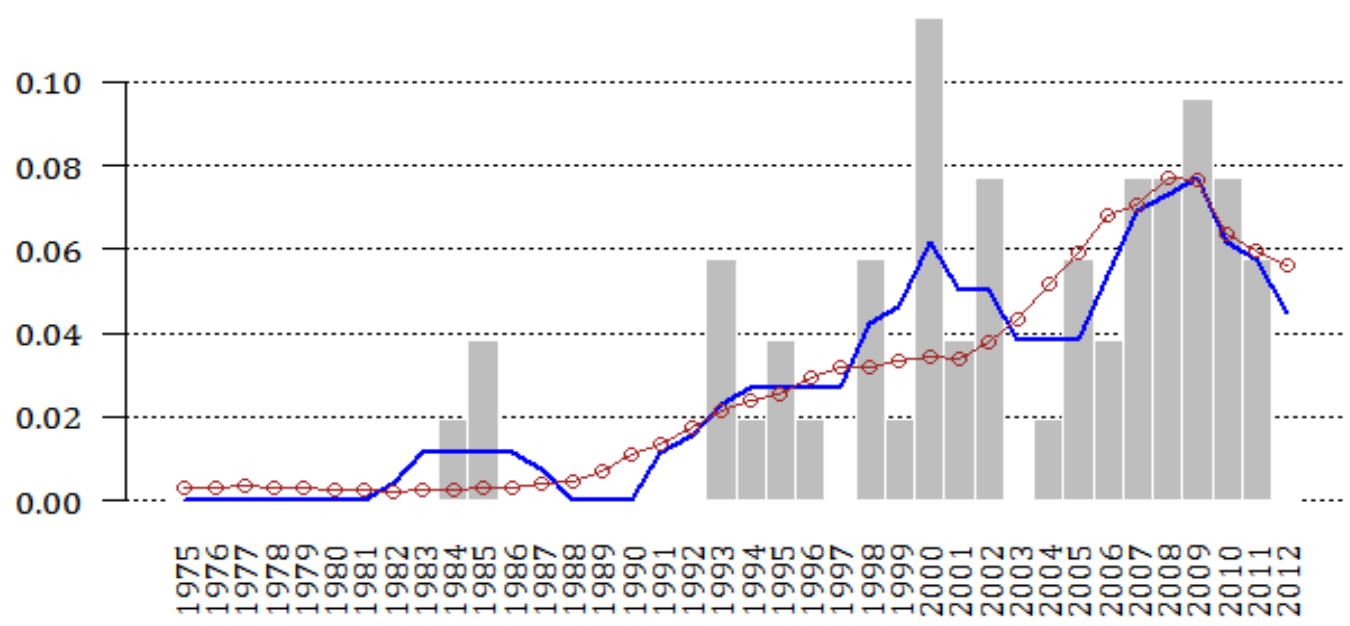

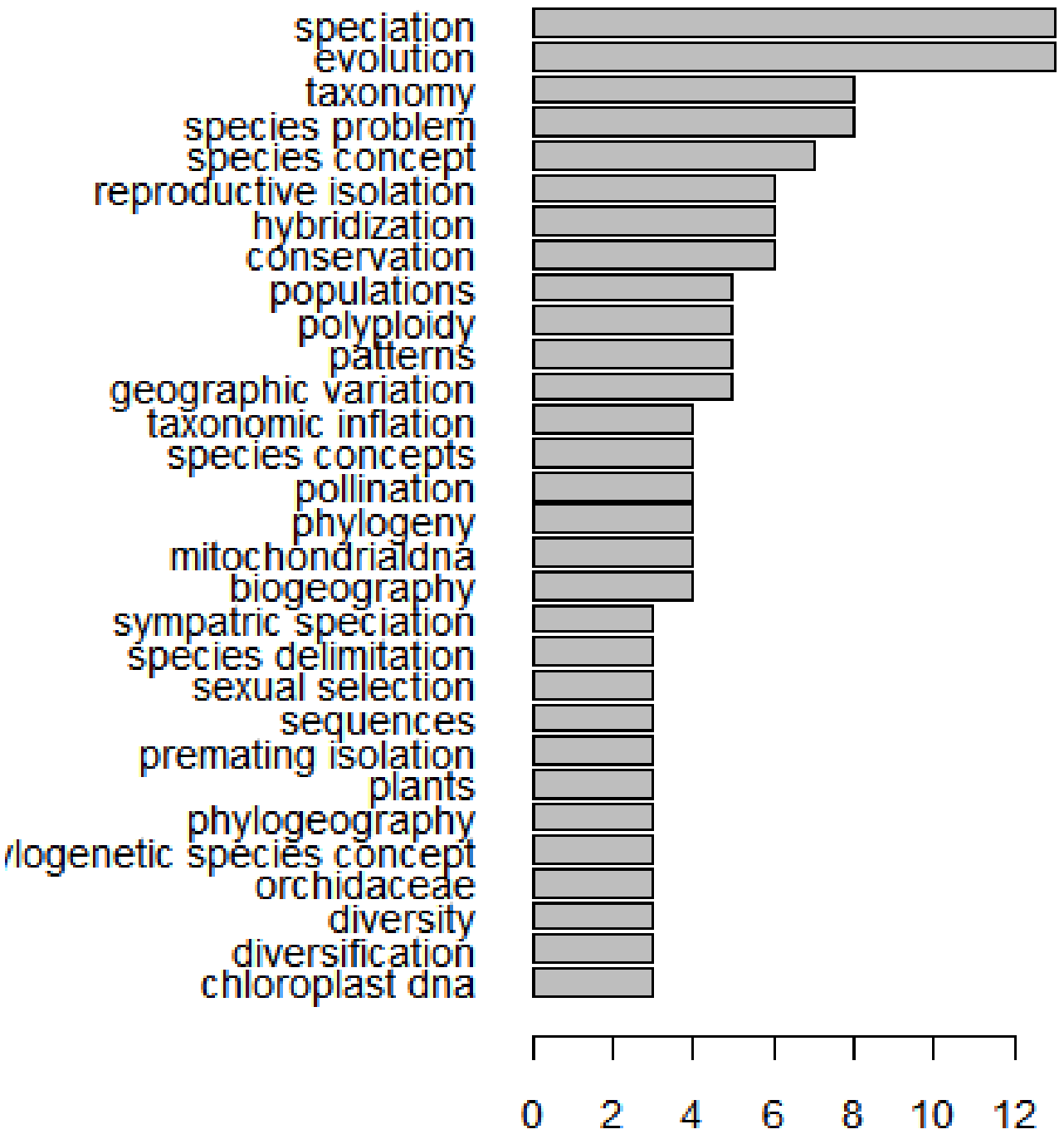

## Cluster 12

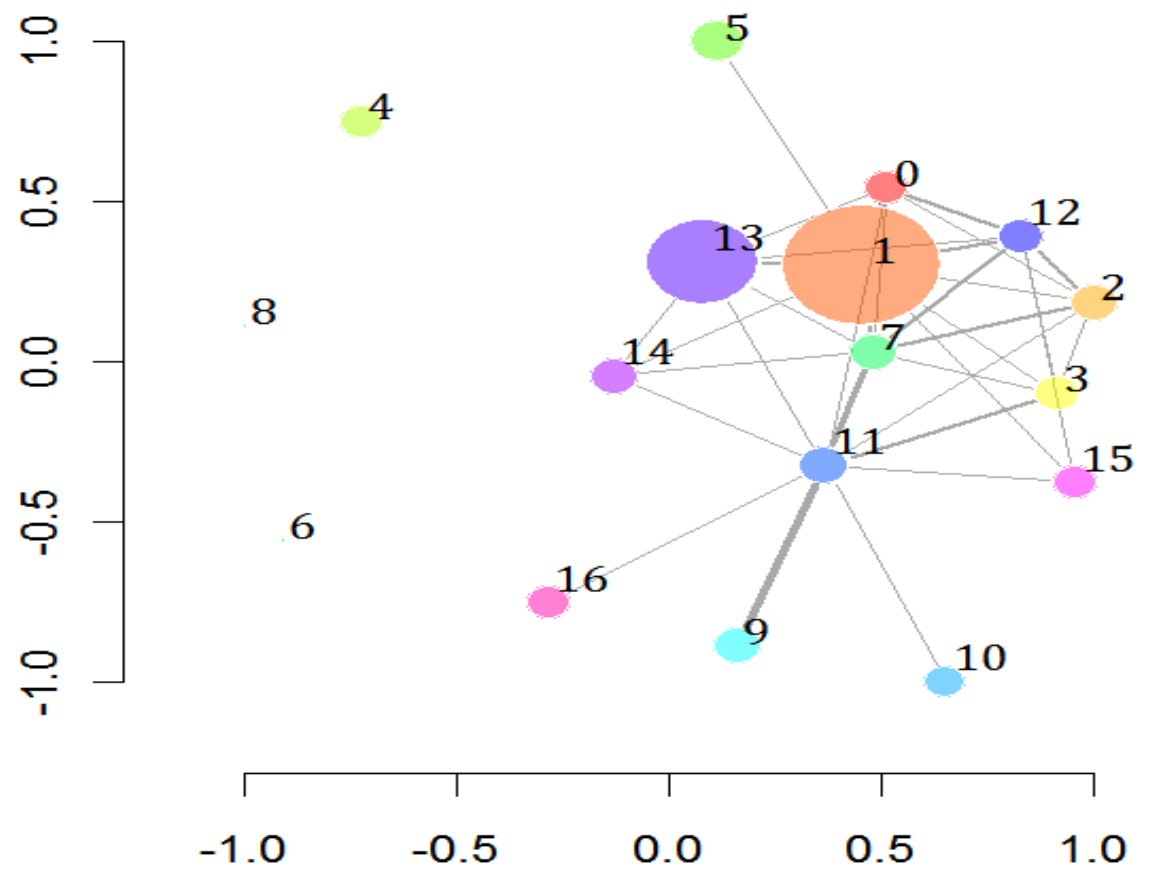

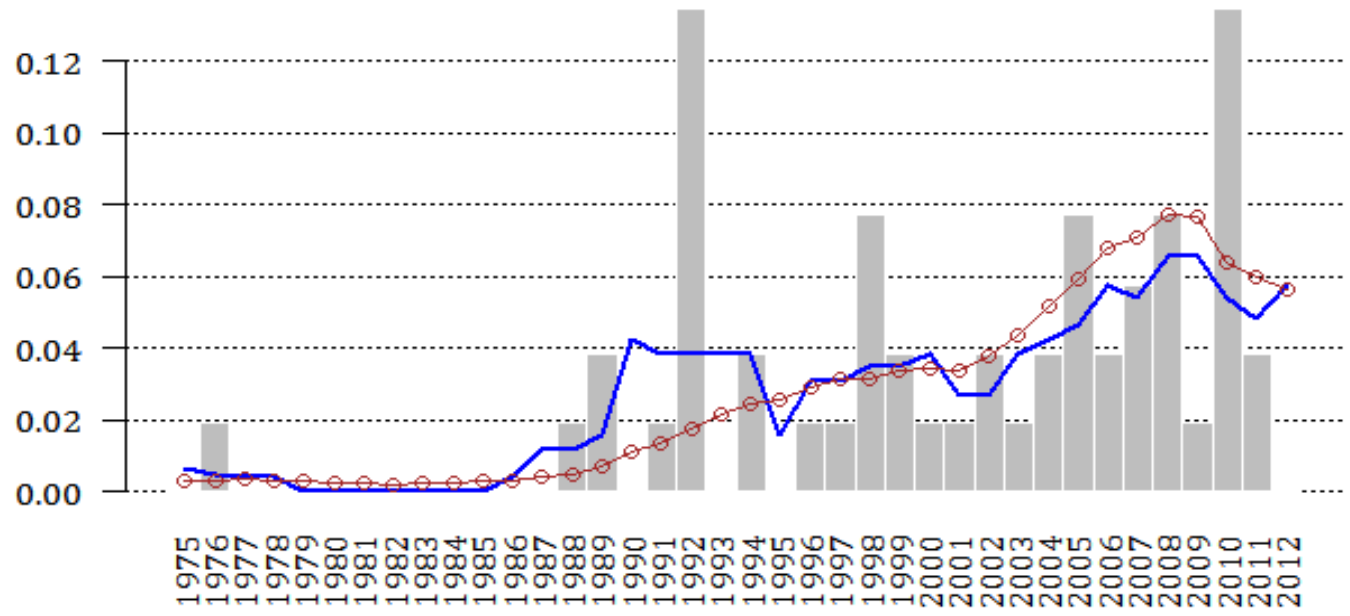

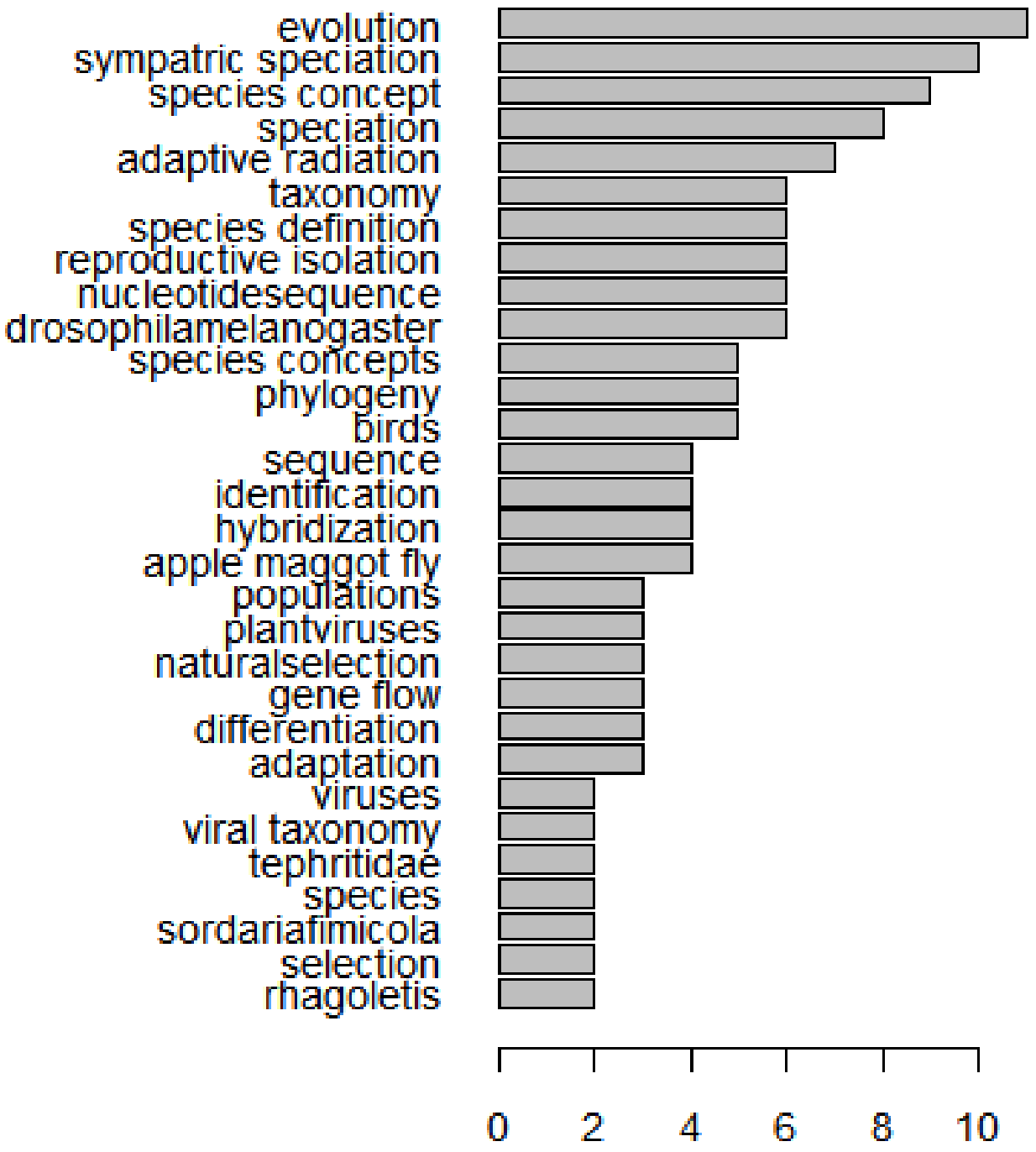

## Cluster 13

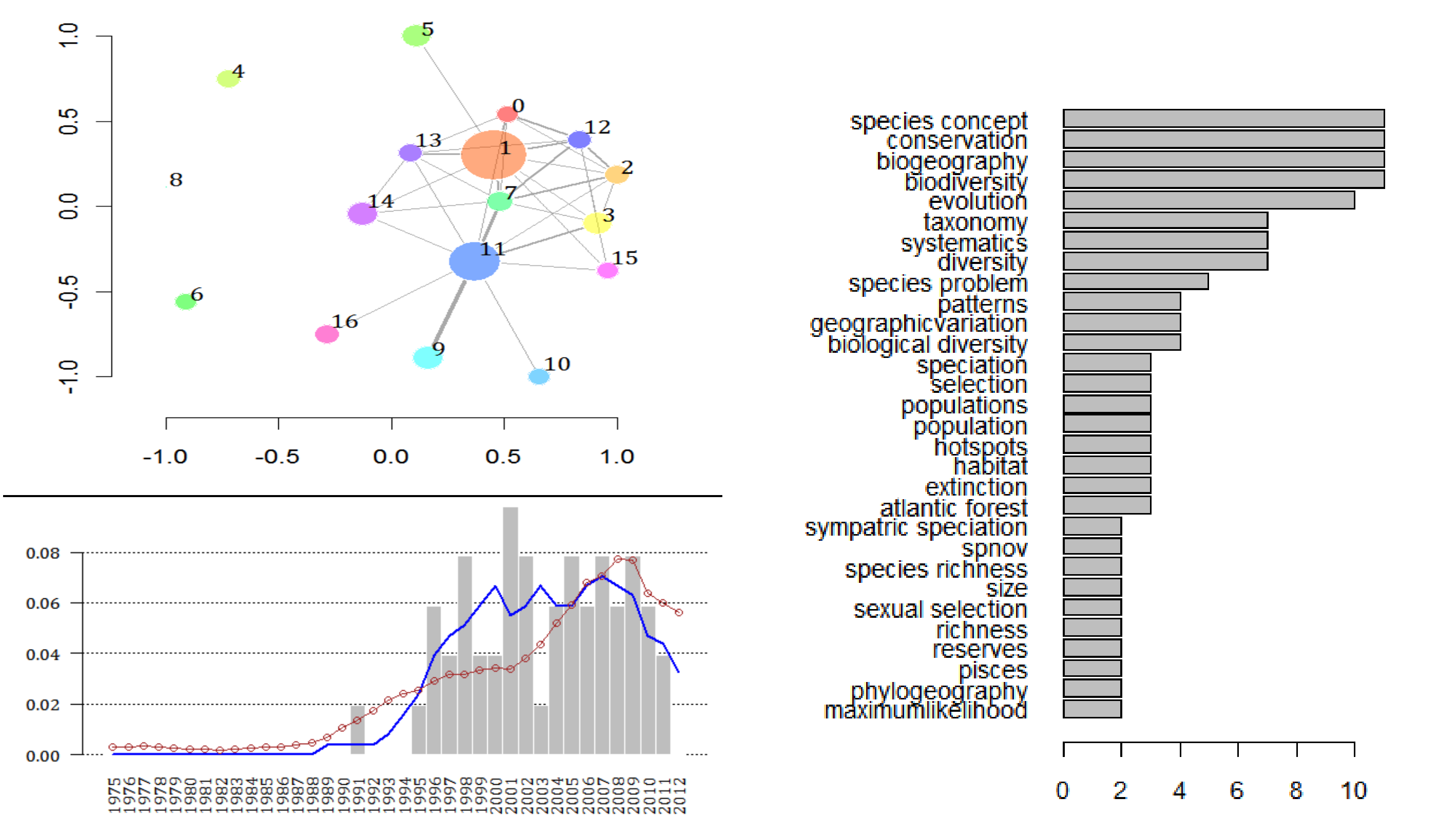

## Cluster 14

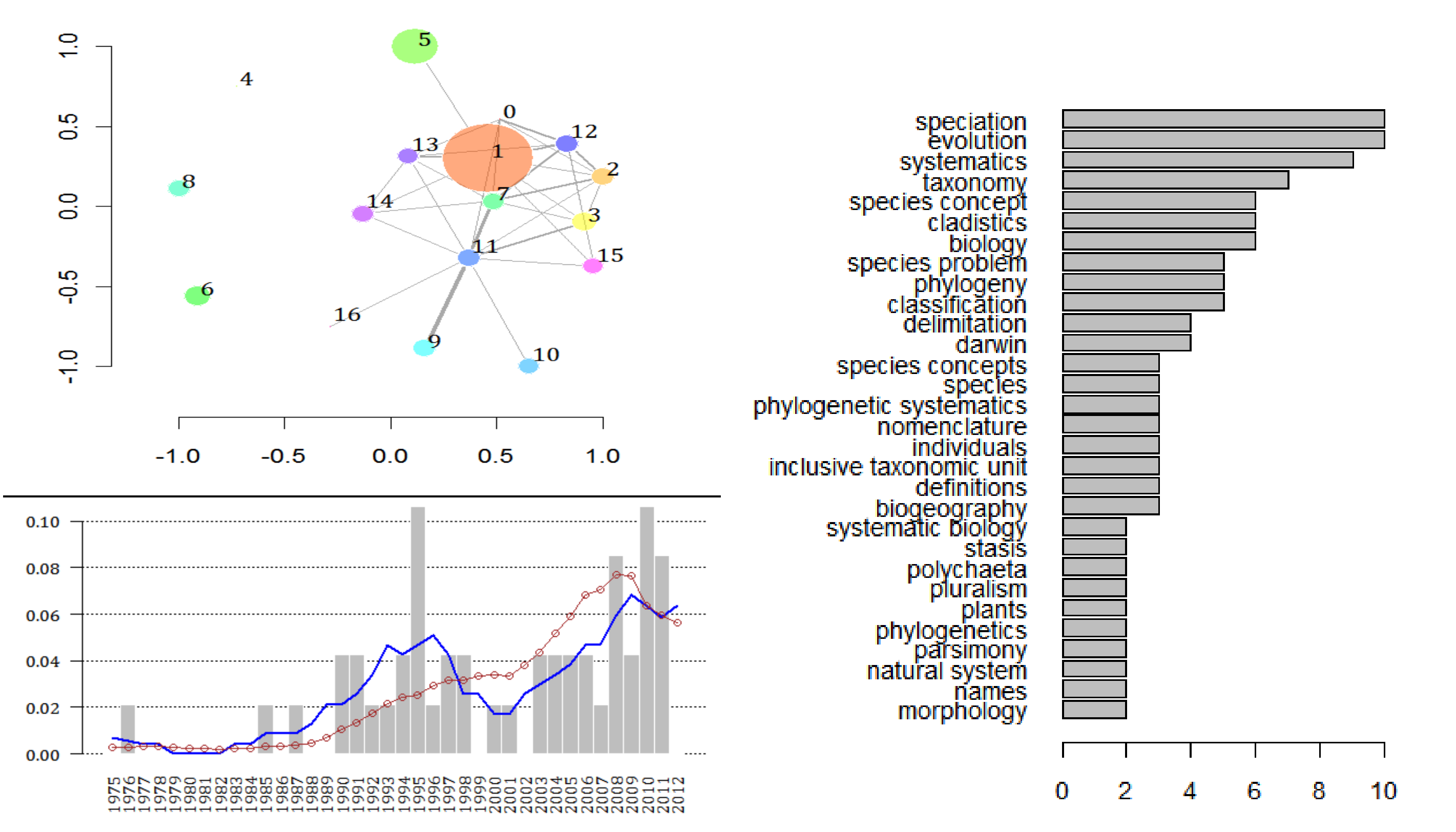

## Cluster 15

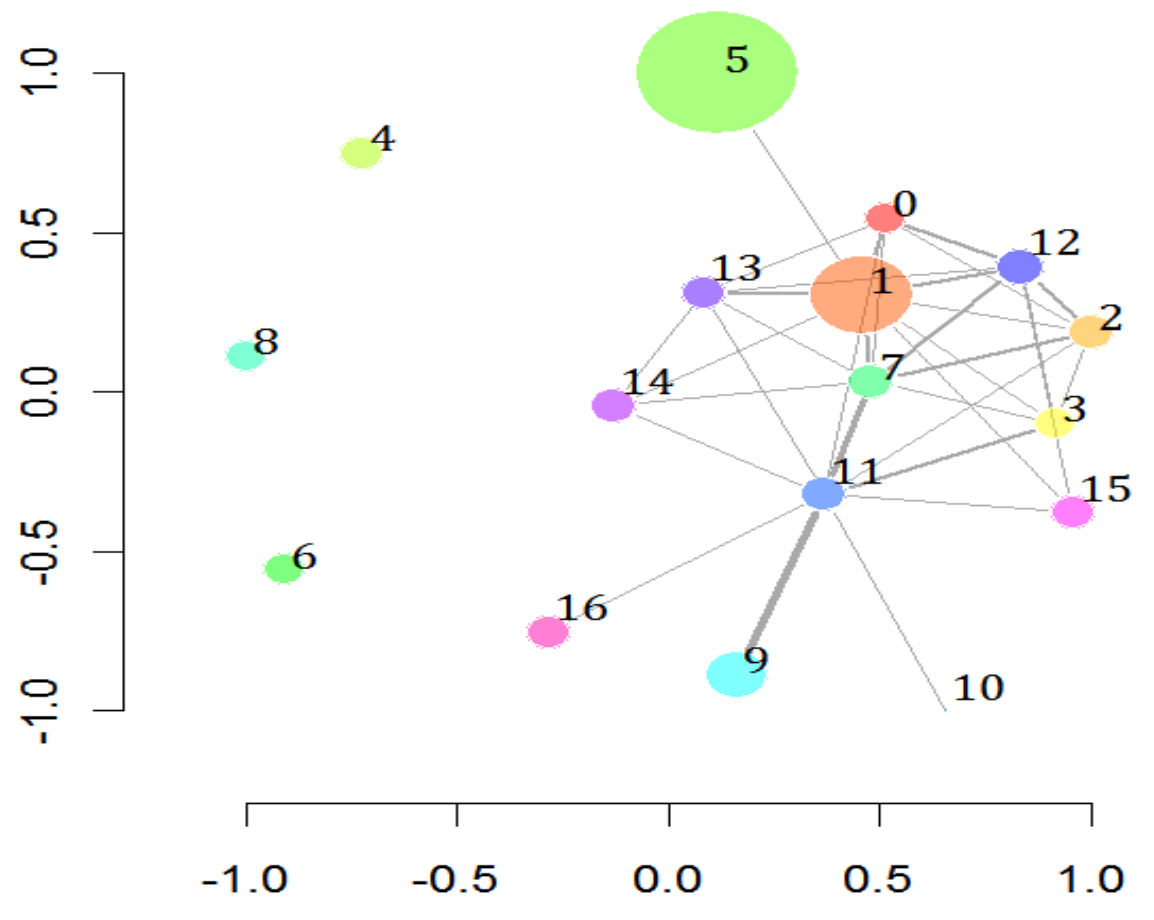

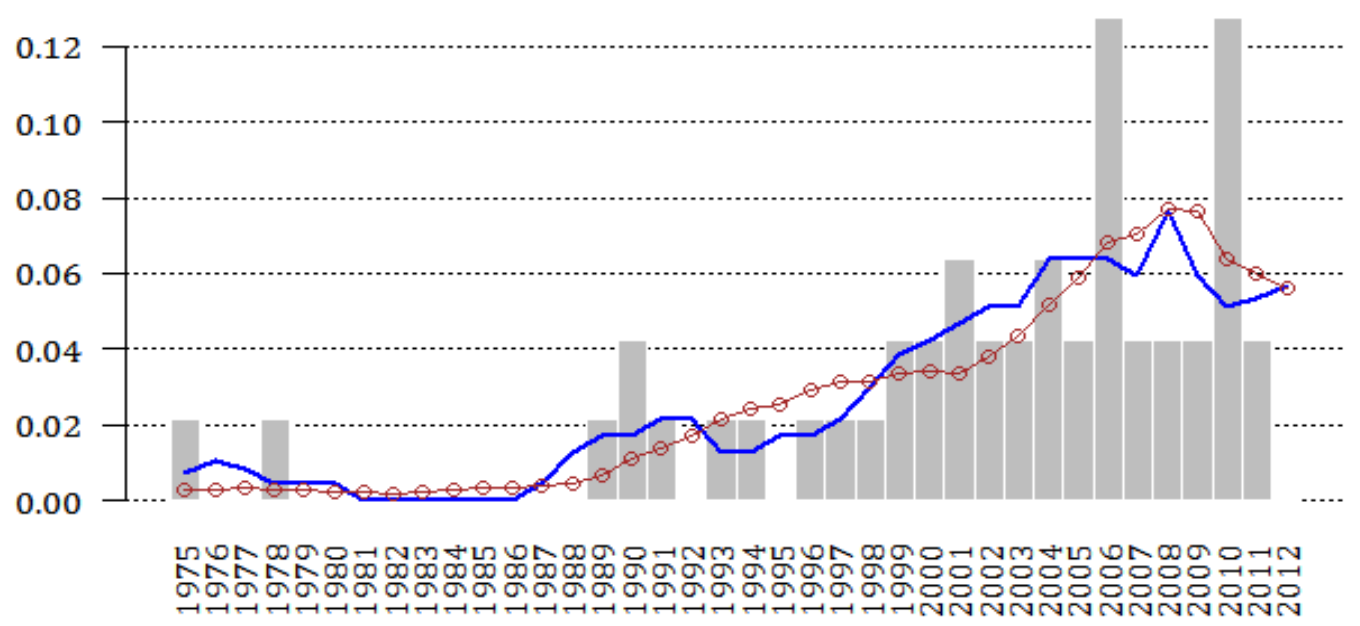

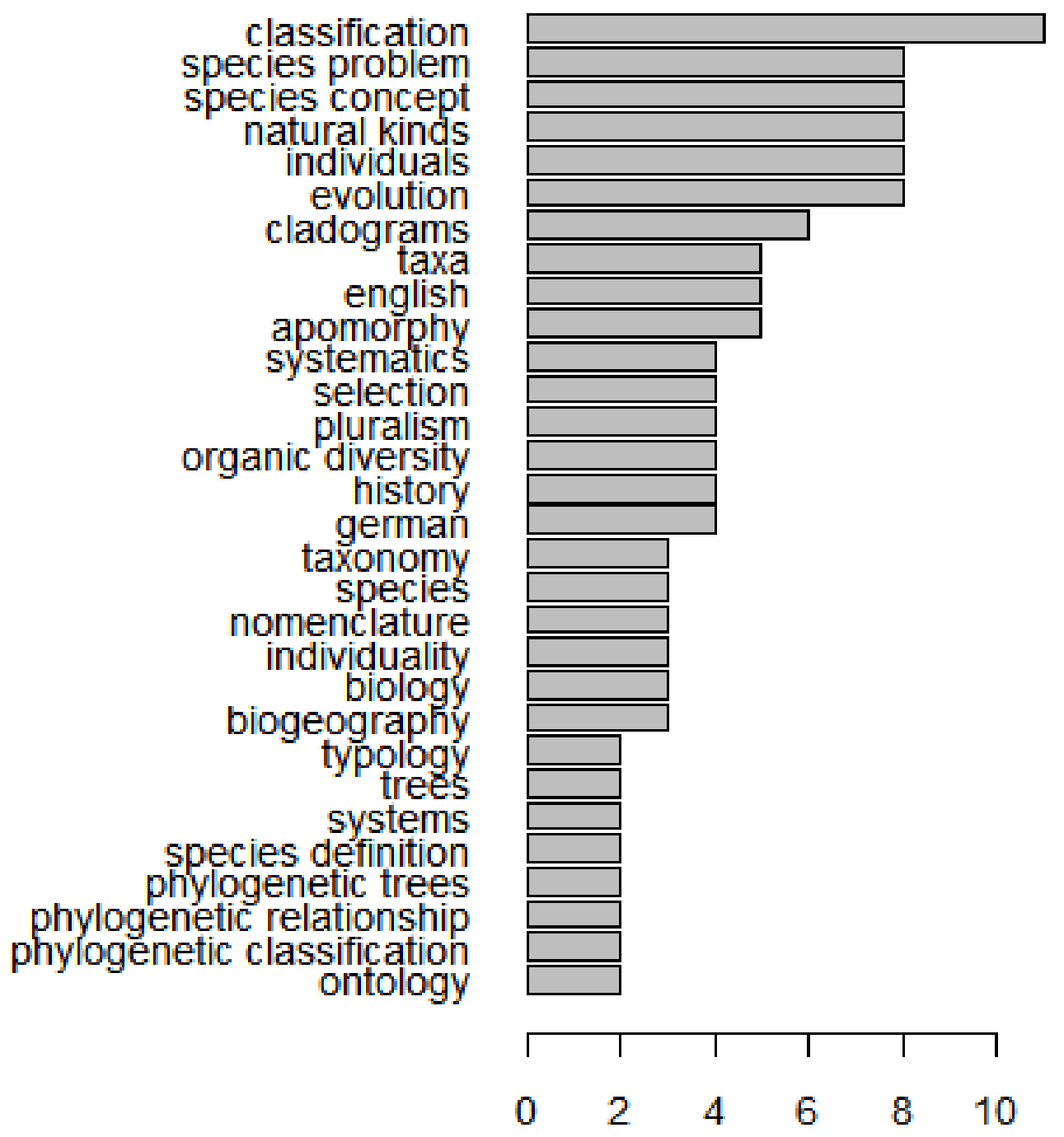

## Cluster 16

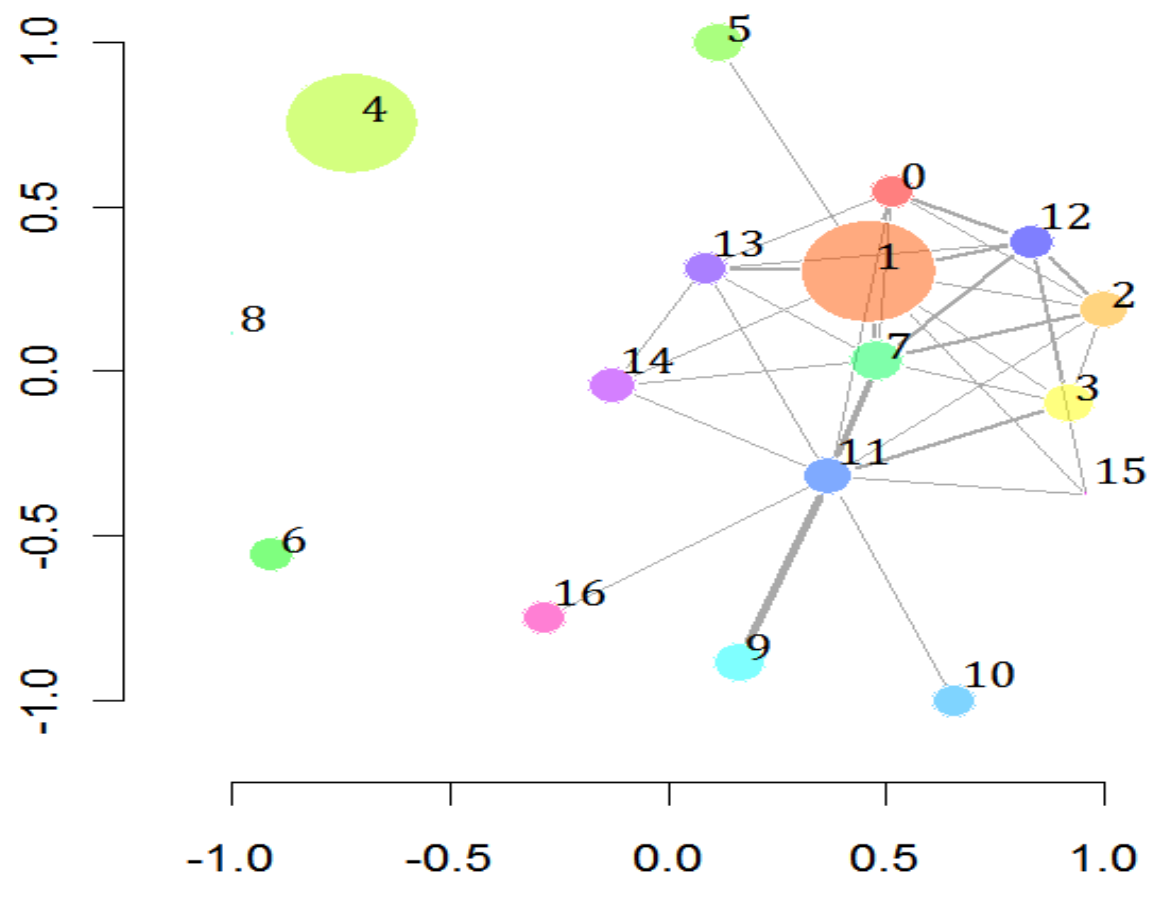

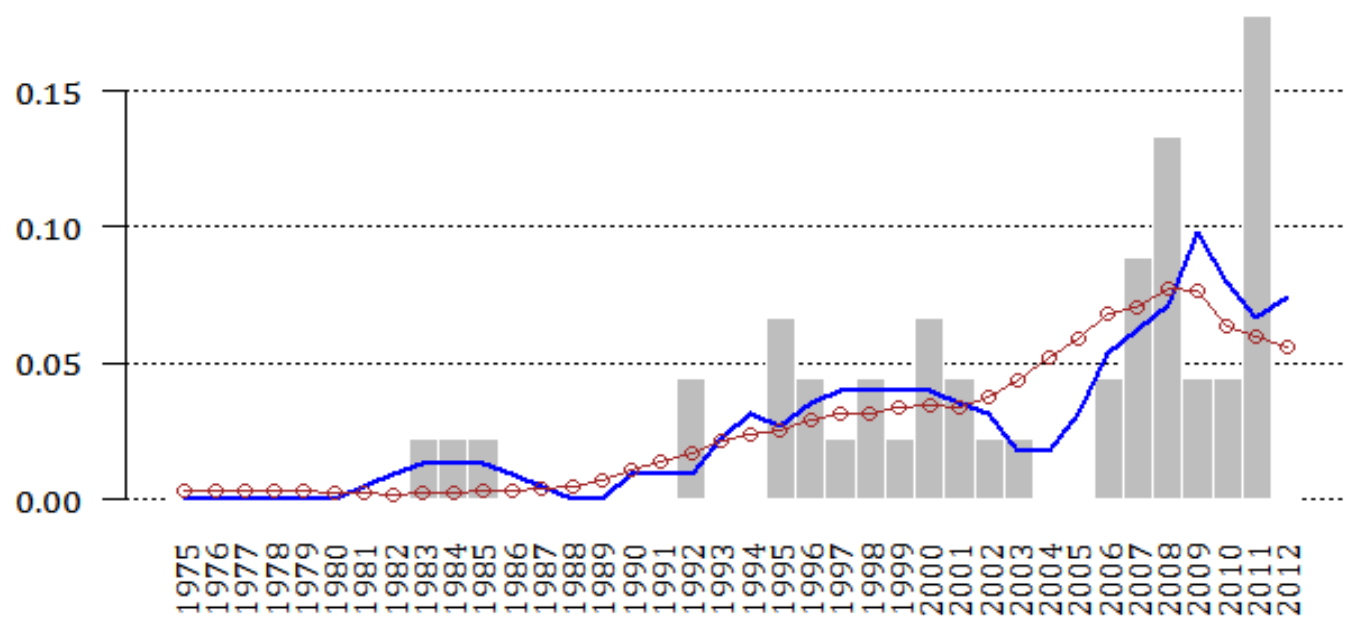

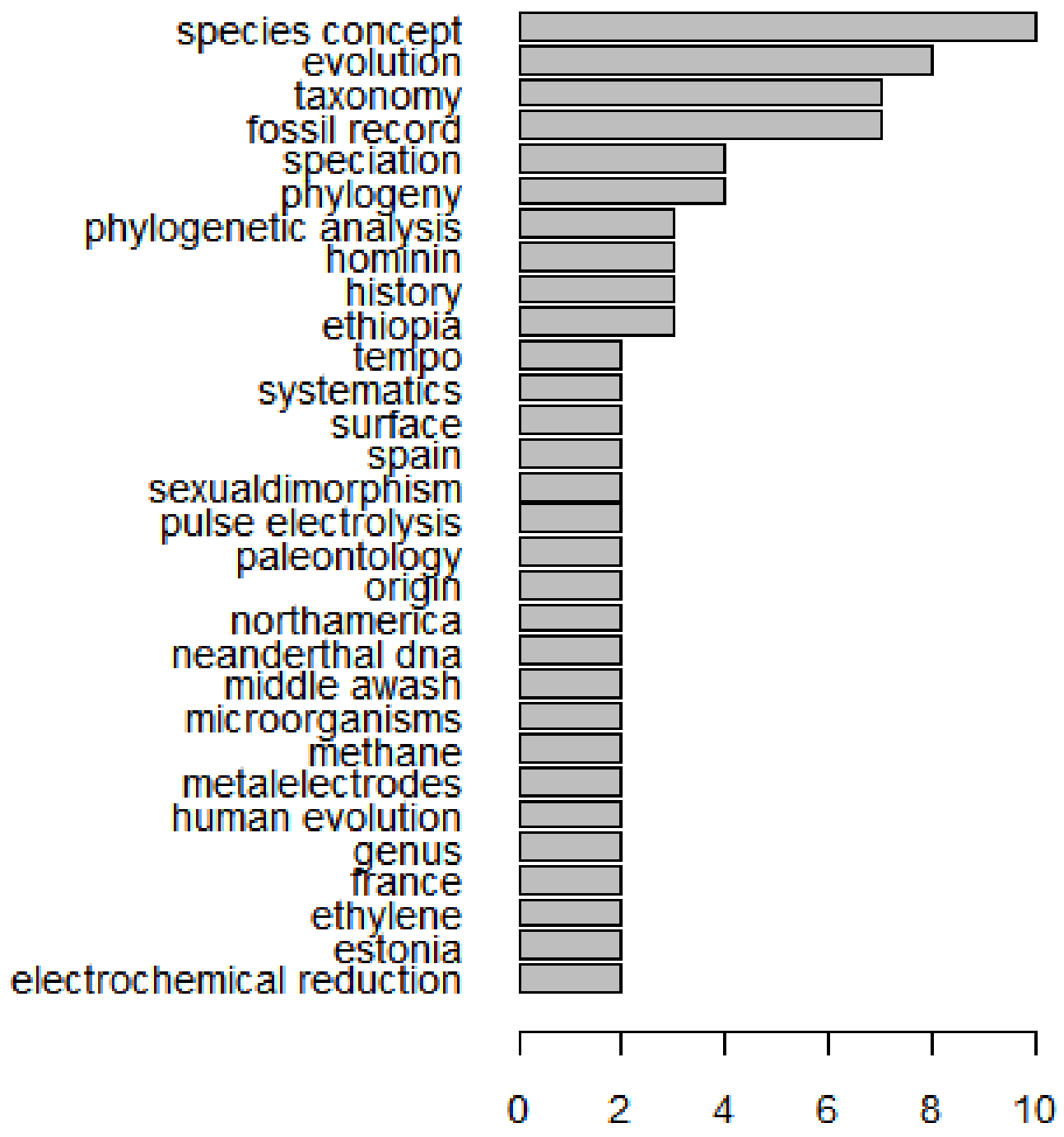

## Cluster 17

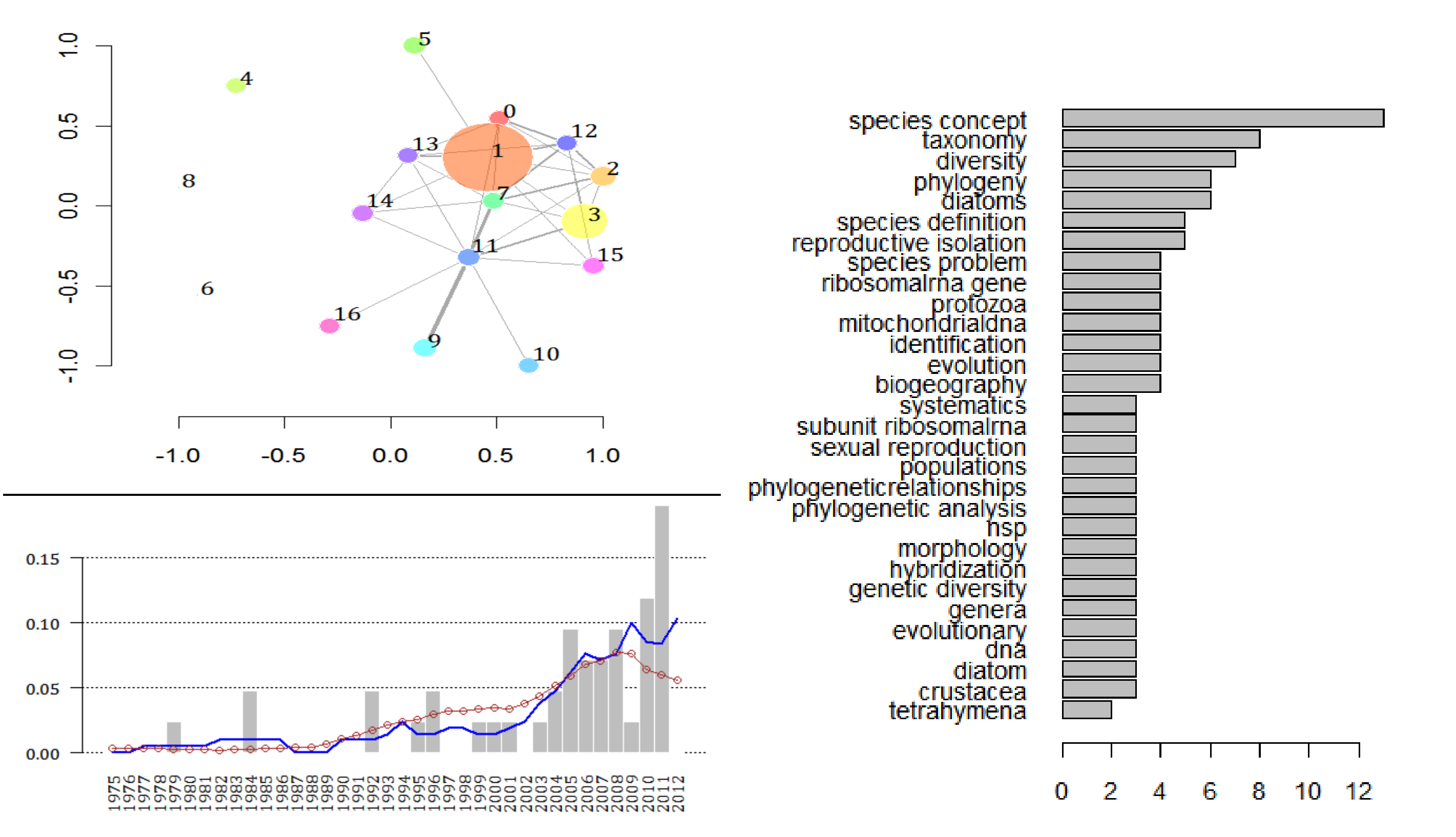

## Cluster 18

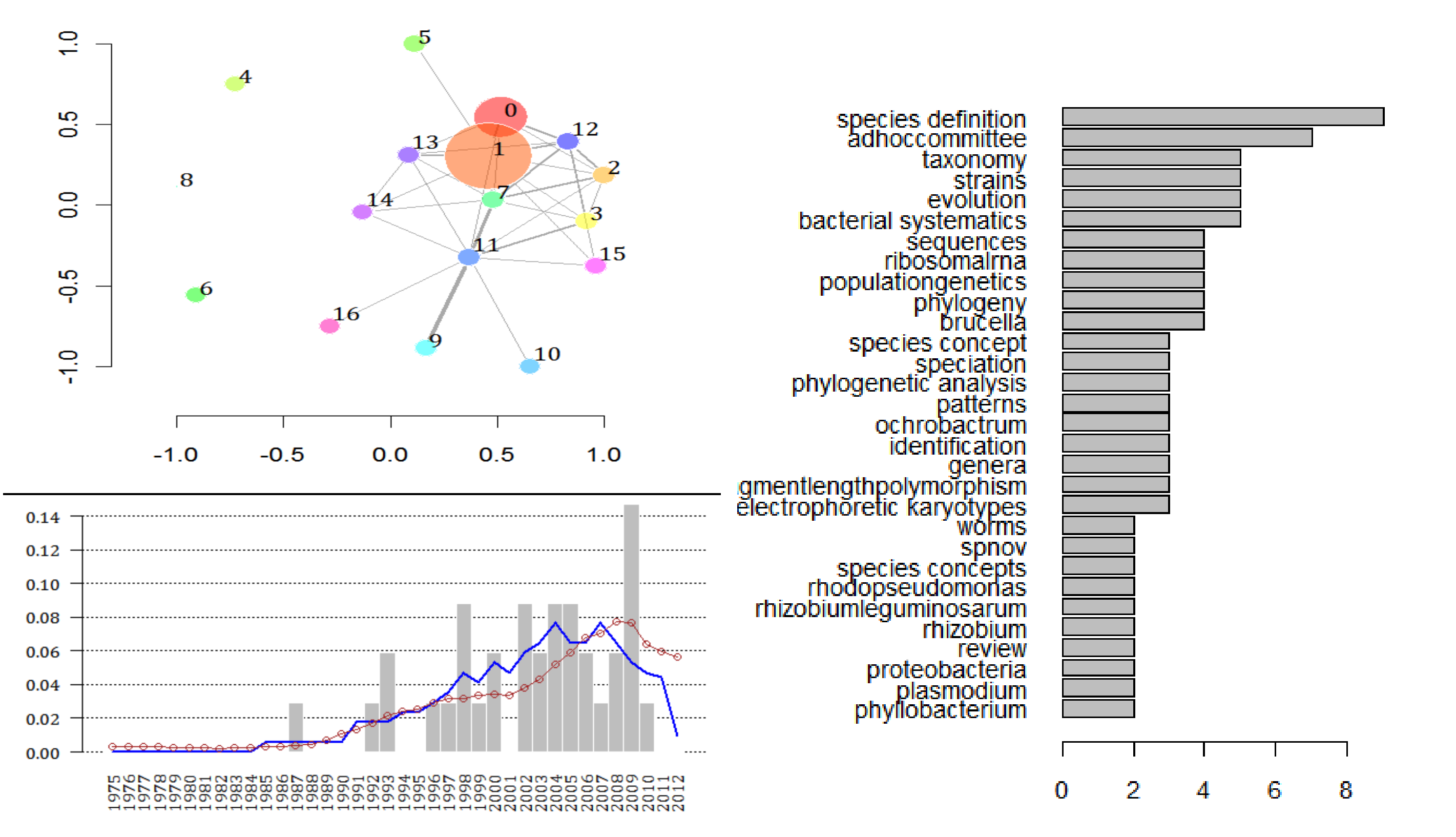

## Cluster 19

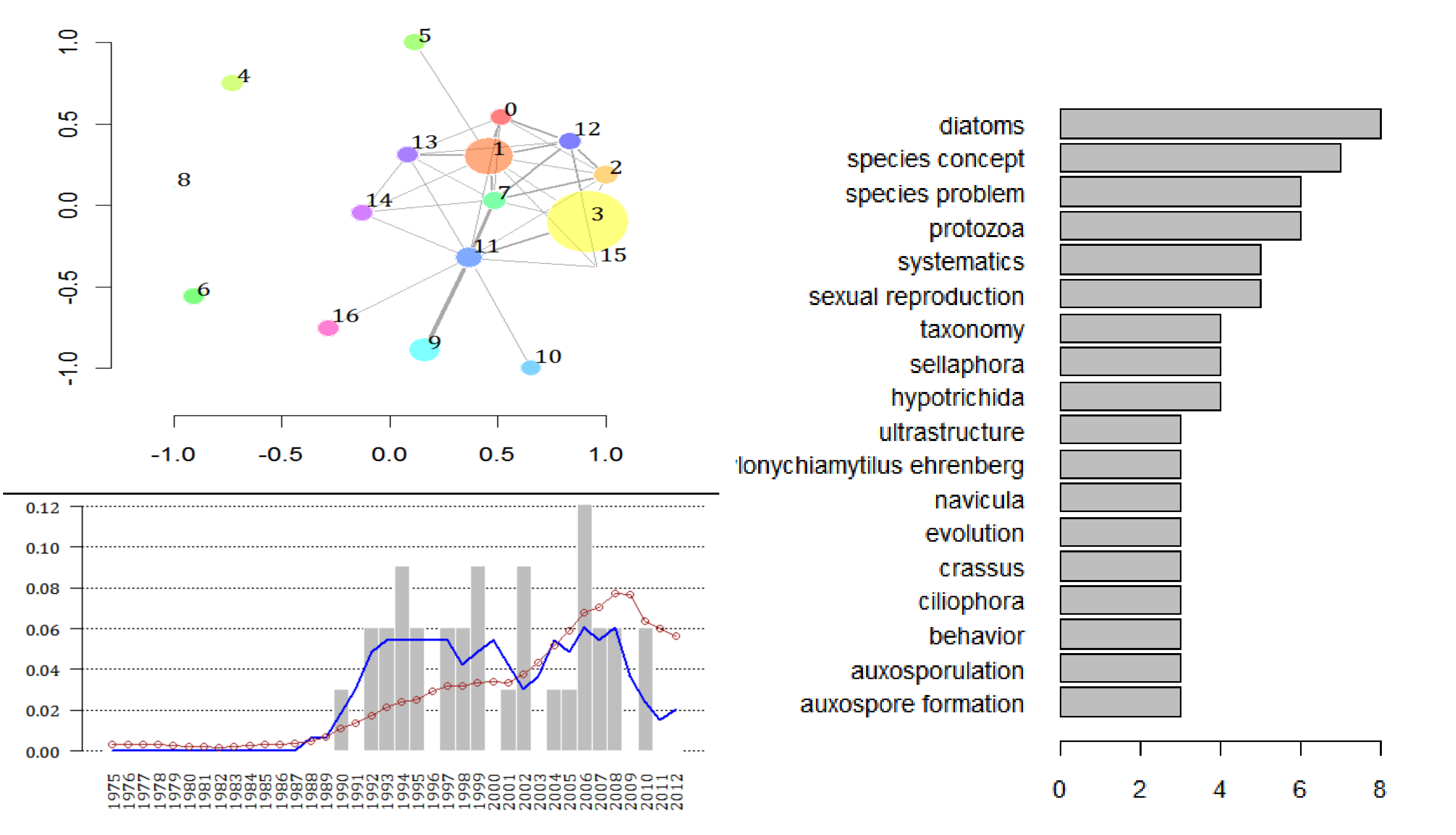

## Cluster 20

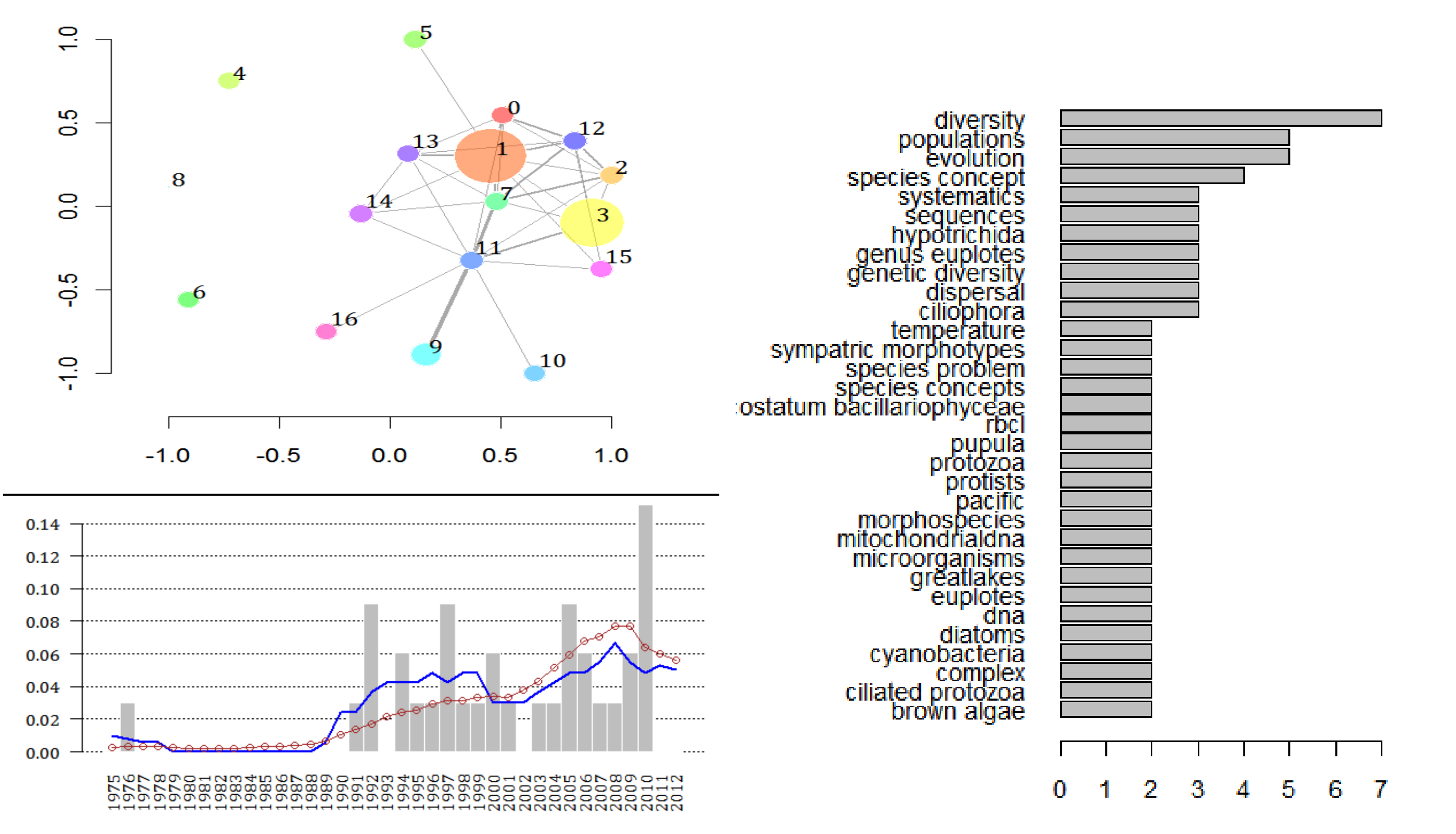

## Cluster 21

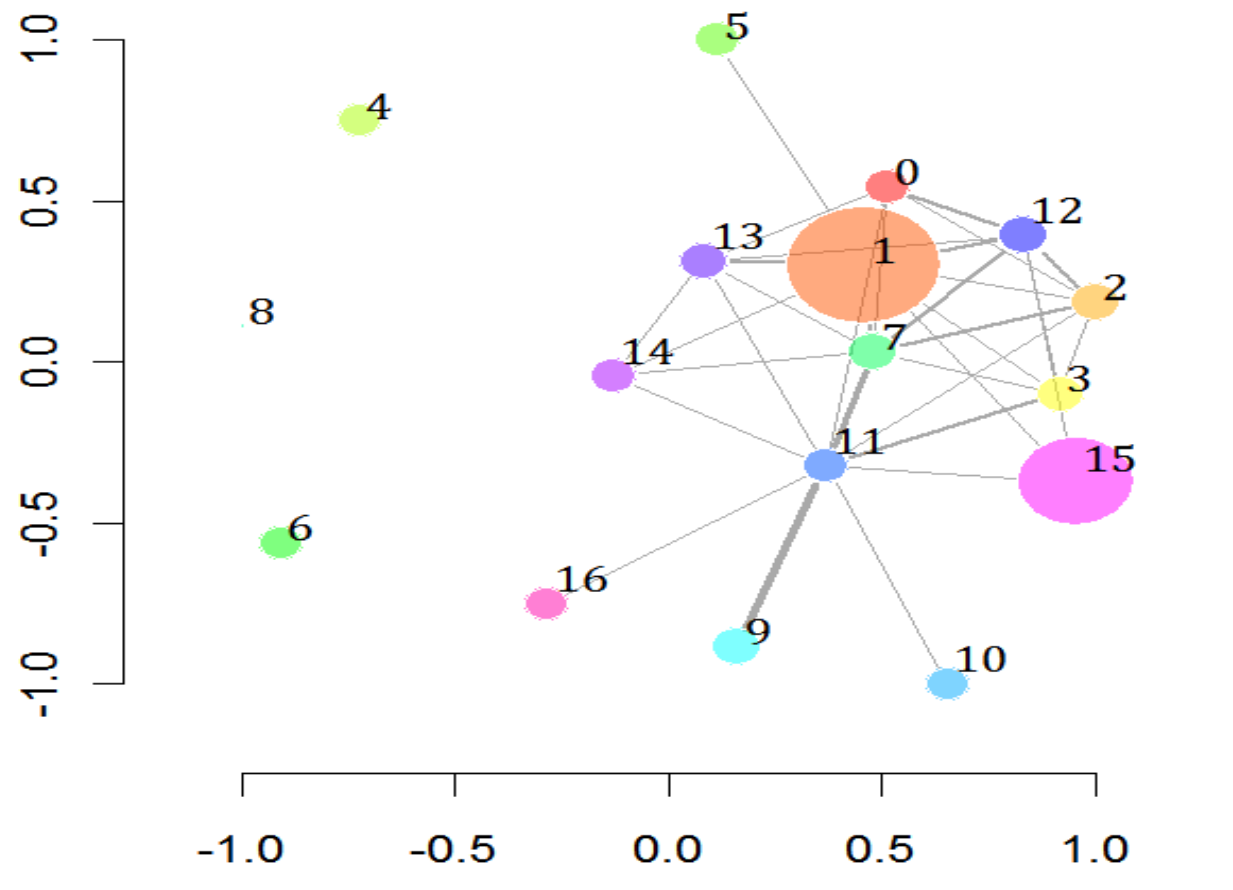

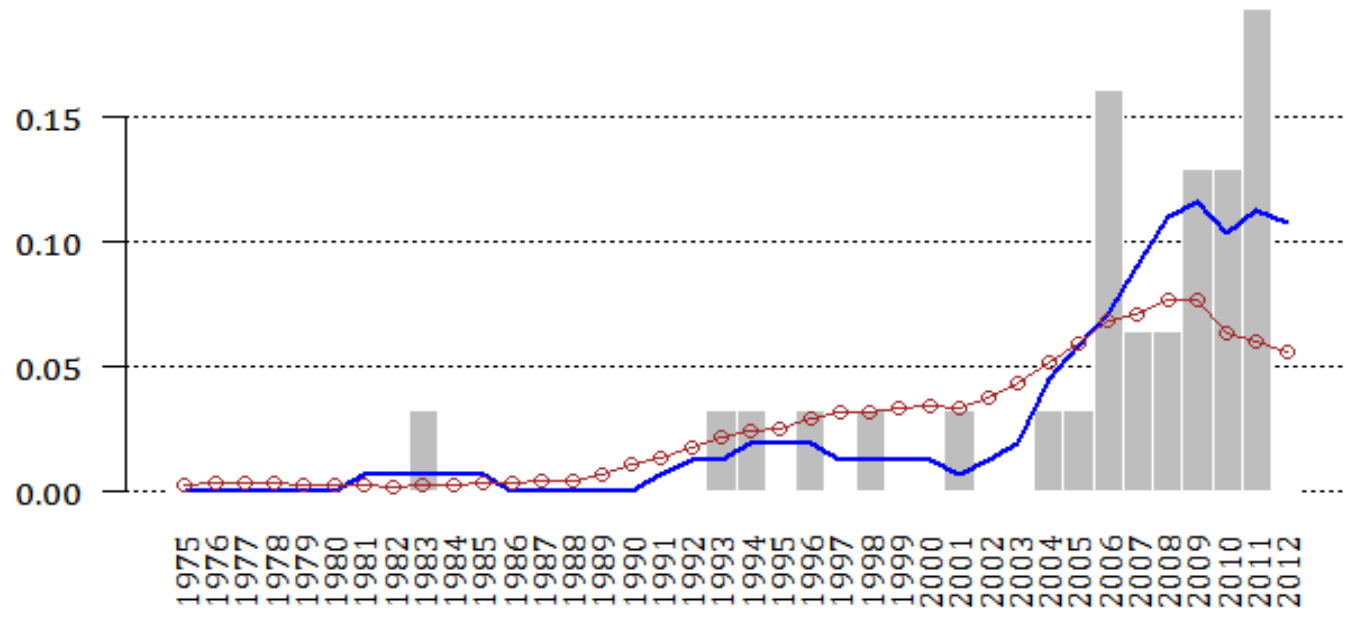

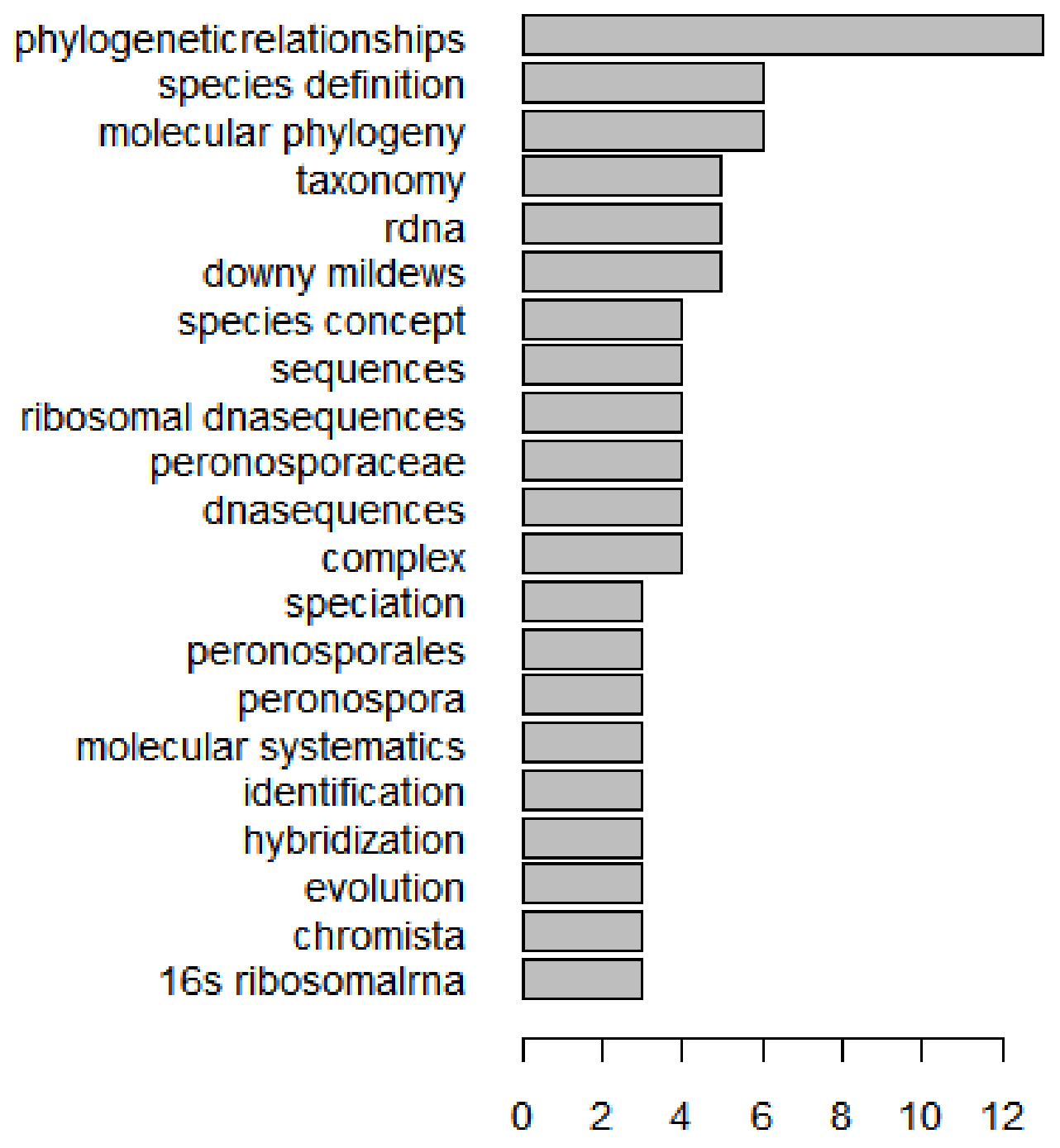